\documentclass[12pt]{article}
\usepackage{amsmath}
\usepackage{graphics}

\newcommand{\rem}[1]{}
\newcommand{\remv}[1]{}

\def\amshow{0}

\newcommand{\onshellflat}{\underset{{\scriptscriptstyle\rm flat }}{\simeq}}
\newcommand{\onshellads}{\underset{{\scriptscriptstyle\rm AdS }}{\simeq}}

\begin{document} 
\begin{titlepage}
\begin{flushright}

\end{flushright}

\begin{center}
{\Large\bf $ $ \\ $ $ \\
Cornering the unphysical vertex 
}\\
\bigskip\bigskip\bigskip
{\large Andrei Mikhailov}
\\
\bigskip\bigskip
{\it Instituto de F\'{i}sica Te\'orica, Universidade Estadual Paulista\\
R. Dr. Bento Teobaldo Ferraz 271, 
Bloco II -- Barra Funda\\
CEP:01140-070 -- S\~{a}o Paulo, Brasil\\
}

\vskip 1cm
\end{center}

\begin{abstract}
In the classical pure spinor worldsheet theory of $AdS_5\times S^5$ there are some 
vertex operators which do not correspond to any physical excitations. We
study their flat space limit. We find that the BRST operator of the
worldsheet theory in flat space-time can be nontrivially deformed without
deforming the worldsheet action. Some of these deformations describe the
linear dilaton background. But the deformation corresponding to the
nonphysical vertex differs from the linear dilaton in not being worldsheet
parity even. The nonphysically deformed worldsheet theory has nonzero
beta-function at one loop. This means that the classical Type IIB SUGRA
backgrounds are not completely characterized by requiring the BRST symmetry
of the classical worldsheet theory; it is also necessary to require the
vanishing of the one-loop beta-function.
\end{abstract}

\end{titlepage}

\tableofcontents

\section{Introduction}
The pure spinor formalism for the classical Type IIB supergravity was 
developed in \cite{Berkovits:2001ue}. As typical for theories with extended supersymmetry, the
formalism is technically challenging and involves many subtle geometrical
constructions. Moreover, even the basic postulates of the formalism are not
completely clear (at least to us). We would like to have some set of axioms
which would allow us to encode the space-time dynamics (SUGRA) in terms of the
worldsheet dynamics. Naively, the set of rules can be as follows:

``Postulate the action of the form:
\begin{align}
\int d\tau^+ d\tau^- \;& \left[ 
   A_{mn}(x,\theta) \partial_+ x^m\partial_- x^n + 
   A_{m\alpha}(x,\theta)\partial_+x^m\partial_-\theta^{\alpha} + \right.
\nonumber \\   
\;&\quad  
+ A_{\alpha m}(x,\theta)\partial_+ \theta^{\alpha} \partial_- x^m + 
A_{\alpha\beta}(x,\theta) \partial_+ \theta^{\alpha} \partial_-\theta^{\beta} \;+
\nonumber \\   
\;&\quad 
+ w_{L+}(\partial_- + A_-(x,\theta)) \lambda_L + w_{R-}(\partial_+ + 
A_+(x,\theta)) \lambda_R \;+ 
\nonumber \\   
\;& \quad\left. + <w_{L+}w_{R-}\lambda_L\lambda_R> \right]
\end{align}
where $\lambda$ and $w$ are pure spinors and their conjugate momenta, and request that
it satisfies the properties:
\begin{itemize}
\item {Classical $2d$ conformal invariance}
\item {Lagrangian is polynomial in $\lambda$ and $w$}
\item {Two separate conserved {\em ghost number} charges, left for 
$\lambda_L\;,\;w_{L+}$ and right for $\lambda_R\;,\;w_{R-}$}
\item {Nilpotent BRST symmetry}
\end{itemize}
The constraints guarantee that these coupling constants $A_{MN}(x,\theta)$ encode a 
solution of the Type IIB SUGRA.''

\vspace{7pt}
\noindent
We believe that this is not very far from the truth, but there are subtleties.

\vspace{7pt}

In order to better understand the pure spinor formalism, it is useful to 
consider explicitly various specific examples beyond the flat space. The most 
symmetric non-flat example is $AdS_5\times S^5$ which was constructed in \cite{Berkovits:2000fe}. 
In \cite{Bedoya:2010qz} we have discussed a special class of deformations of $AdS_5\times S^5$ known 
as $\beta$-deformations. At the linearized level, we have explicitly constructed 
the corresponding 
deformations of the pure spinor action. They are described by the integrated 
vertex operators, which are products of two global symmetry currents with
some constant coupling constant $B^{ab}$:
\begin{equation}\label{GeneralBeta}
U = B^{ab} j_a\wedge j_b
\end{equation}

\subsection{Non-physical vertices}
As was pointed out in \cite{Bedoya:2010qz}, some apparently well-defined vertex operators of 
the form (\ref{GeneralBeta}) do not correspond to any physical deformations of the
$AdS_5\times S^5$ background. They have: 
\begin{equation}
B^{ab}f_{ab}{}^c \neq 0
\end{equation}
where $f_{ab}{}^c$ is the structure constants of the SUSY algebra ${\bf g} = {\bf psu}(2,2|4)$.
We will call such vertices ``non-physical''. Their appearence does not lead 
to any obvious contradiction, if one can either consistently throw them away, 
or perhaps learn to live with them. Throwing them away should presumably 
correspond to an additional restriction of the allowed BRST cochains, similar 
to the semi-relative cohomology of the bosonic string \cite{Nelson:1988ic,Distler:1990ea,Witten:1992yj}. 

In this paper we will study the flat space limit of these unphysical vertices.

\subsection{Flat space limit of SUGRA excitations}
We will start by pointing out  the following general fact about the
flat space limit of SUGRA solutions.

Given a general nonlinear solution (``the background'') of the Type IIB SUGRA 
we can consider the linear space of its infinitesimal deformations 
(``excitations''). Such excitations correspond to solutions of certain 
{\em linear} differential equations, namely the SUGRA equations of motion 
linearized around this background). 

In particular, let us look at the flat space limit of the excitations of
$AdS_5\times S^5$. Both the flat space sigma-model and the sigma-model of 
$AdS_5\times S^5$ are invariant under a parity symmetry. Therefore linearized
excitations can be separated into parity-odd excitations and parity-even 
excitations. Let us restrict ourselves to the bosonic excitations. Those
excitations which involve NSNS and RR $B$-fields ({\it i.e.} RR 3-form field
strength) are parity-odd, while those which involve metric, dilaton, axion,
and the RR 5-form field strength are parity-even. 

Let us pick some particular excitation and look at its Taylor expansion 
around a fixed ``marked point'' $x_*$. Consider {\em only the leading term} in 
the Taylor expansion. It is a polynomial in $x-x_*$. We claim that:
\begin{equation}\label{StatementOnLeadingTerm}
\begin{array}{l}
\mbox{The leading term of a parity-odd excitation of $AdS_5\times S^5$} 
\cr
\mbox{is a polynomial solution of the flat space linearized SUGRA}   
\end{array}
\end{equation}
\paragraph     {Proof} 
Equations of motion of Type IIB SUGRA are systematically reviewed in \cite{Green:1987mn}.
For the leading approximation to the flat space limit of $AdS_5\times S^5$, we get
the following linearized equations for $H_{NSNS}$ and $H_{RR}$:
\begin{align}\label{EqMH}
   d * (H_{NSNS} + iH_{RR}) = -{2\over 3} i * \iota_{(H_{NSNS} + iH_{RR})} F_5
\end{align}
where $\iota_{(H_{NSNS} + iH_{RR})} F_5$ is the substitution of the complex 3-form
$H_{NSNS} + iH_{RR}$ into the RR 5-form field strength of $AdS_5\times S^5$.
We have to prove that the leading term of $H_{NSNS} + iH_{RR}$ is a solution of 
the linearized SUGRA in flat space. We expand (\ref{EqMH}) in Taylor series.
For the term with the leading power of $x$, all that matters is the term with
the maximal number of derivatives. It is the same as in flat space:
\begin{equation}
   d *_{\rm flat} (H_{NSNS} + iH_{RR}) = 0
\end{equation}

\subsection{Flat space limit of non-physical vertices}
Although the non-physical vertices do deform the AdS action consistently, 
and in
a BRST-invariant way, they do not correspond to any linearized supergravity 
solution. We can see it in the flat space limit. We expand the vertex around
a fixed ``marked point'' $x_* \in AdS_5\times S^5$ and look at the leading term.
We observe that the SUGRA fields read from the leading term do not solve the 
linearized SUGRA equations in flat space.  This confirms the observation 
of \cite{Bedoya:2010qz} that the non-physical vertex does not correspond to any deformation of 
$AdS_5\times S^5$. If the non-physical vertex corresponded to a valid deformation
of $AdS_5\times S^5$, then this would be in contradiction with (\ref{StatementOnLeadingTerm}).

Moreover, it turns out that 
there is an essential difference between the non-physicalness of the AdS 
deformation vertex (\ref{GeneralBeta}) and its flat space limit. In case of AdS, the vertex 
given by Eq. (\ref{GeneralBeta}) at least deforms the worldsheet action in $AdS_5\times S^5$ in 
a consistent way. Its flat space limit, however, does not even provide a 
consistent deformation of the flat space worldsheet action. How can it be?

\subsection{Wild deformations of the BRST operator}\label{sec:WildDeformations}
The mechanism is the following. Remember that usually the BRST-invariant 
deformations of the worldsheet action are accompanied by the corresponding 
deformation of the BRST operator\footnote{because the BRST-invariant integrated vertex is only 
BRST-invariant on-shell} $Q$. The deformations of the BRST 
structure are tied to the deformations of the action. But in the special case 
of flat space there are ``wild''
deformations of the BRST structure, which do not require the deformations
of the action: 
\begin{itemize}
\item We can deform the BRST structure keeping the action fixed. 
\end{itemize}
We will call these deformations of $Q$ ``wild'', in the sense that they are 
not tied to the deformations of the action.
These ``wild'' deformations of the BRST structure play an important role in 
the flat space limit of the unphysical $\beta$-deformations. Let us consider a 
$\beta$-deformation of the AdS space and expand everything around flat space. 
If the expansion of the $\beta$-deformation vertex starts from $R^{-3}$, then
the flat space limit is perfectly physical; it is just a constant RR 3-form
field strength. But for some vertices (or, equivalently, for some
choice of the expansion point $x_*\in AdS_5\times S^5$) the expansion starts 
with $R^{-4}$. In this case we get\footnote{Usually the action is defined with the overall coefficient $R^2$; then the flat space term is of the order $1$. We prefer to define the action so that the flat space is of the order $R^{-2}$.}:
\begin{equation}
S = R^{-2}S_{\rm flat} + \int R^{-3}U_{\rm AdS\; RR\; 5-form} + \int \varepsilon R^{-4}U_{\beta} + \ldots
\label{OrderExpansionOfS}
\end{equation}
Here $R$ is the radius of AdS space, $\varepsilon$ the small parameter measuring the 
strength of the $\beta$-deformation,  $U_{\rm AdS\; RR\; 5-form}$ is the integrated vertex 
corresponding to the deformation of flat space into AdS, and $U_{\beta}$ is the 
leading term in the expansion of the $\beta$-deformation integrated vertex around 
the marked point. It turns out that the BRST operator of the unphysical 
$\beta$-deformation, in the flat space expansion, contains a {\em wild piece} at
the lower order then one would expect:
\begin{equation}
   Q = Q_{\rm flat} + \varepsilon R^{-1}\Delta_{\rm wild}Q + \ldots
\end{equation}
where $\Delta_{\rm wild}Q$ is a wild deformation of $Q_{\rm flat}$. Note that the BRST operator
gets deformed at the order $R^{-1}$, although naively one would expect $R^{-2}$.
Then we get:
\begin{align}
(\Delta_{\rm wild}Q)\; S_{\rm flat} = & \; 0
\\     
(\Delta_{\rm wild}Q)\; U_{\rm AdS\; RR\; 5-form} = &\; Q_{\rm flat} \;U_{\beta}
\label{DefUB}
\end{align}
This means that $U_{\beta}$ is not even BRST closed. 

In other words, when studying the flat space limit of this $\beta$-deformation, 
it only makes sense to consider the deforming vertex up to the relative 
order $R^{-1}$. But as we see in Eq. (\ref{OrderExpansionOfS}), the beta-deformation starts only at 
the relative order $R^{-2}$ (the term with  $U_{\beta}$). In this sense, the flat space
limit of our beta-deformation only affects the BRST operator {\em without 
touching the action}.

\subsection{Deformations of the normal form of the action}

However, as explained in \cite{Berkovits:2001ue}, in order to read the SUGRA fields from the
worldsheet action, we have to first bring the action to some special 
{\em normal form}. The definition of this normal form does depend on the BRST 
operator; therefore the {\em normal form} of the action does get deformed in 
the flat space limit. We will discuss this in Section \ref{sec:NormalForm}. We will find that 
the leading term in the near-flat space expansion of the nonphysical vertex 
would have resembled the linear dilaton, but differs from it in not being 
worldsheet parity invariant. This leads to the axial asymmetry of the vector 
components of the worldsheet Weyl connection, and consequently to the  
{\bf anomaly} at the one-loop level. 

\paragraph     {Conclusion:} 
A classical Type IIB background is not completely characterized by requiring 
the BRST symmetry of the classical worldsheet theory; it is also necessary to 
require the vanishing of the one-loop beta-function.  

\paragraph     {Open question:}
It is not clear to us if there exists such nonphysical vertices in the
backgrounds other than flat space and $AdS_5\times S^5$. We suspect that, even
forgetting about the quantum anomaly, the non-physical deformation of the 
classical sigma-model will be obstructed at the higher orders of the 
deformation parameter. 

\vspace{10pt}
\noindent In the rest of the paper we will provide technical details.

\section{$AdS_5\times S^5$ and its $\beta$-deformations}
\subsection{Pure spinor formalism in $AdS_5\times S^5$}
\subsubsection{The action}
The action is:
\begin{align}\label{SAdS}
S_{\rm AdS} =&\;    
\int d^2 z\, \hbox{Str} \left( {1\over 2} J_{\bar{2}+}J_{\bar{2}-} +
{3\over 4} J_{{\bar{1}}+}J_{\bar{3}-} +{1\over 4} J_{\bar{3}+}J_{\bar{1}-}  
+ \mbox{ \tt [ghosts]}  \right)
\end{align}
where the currents are $J = -dg g^{-1}$, $g = e^{\theta}e^x$, and the indices with the bar
denote the ${\bf Z}_4$ grading. 

\subsubsection{Parity symmetry}
There is a parity symmetry $\Sigma$:
\begin{align}
\Sigma(\tau^{\pm}) =\;& \tau^{\mp}
\nonumber \\    
\Sigma(g) = \;& SgS^{-1}
\label{AdSParity}
\end{align}
where $S$ is an element of $PSU(2,2|4)$ given by the following
$(4|4)\times (4|4)$-matrix:
\begin{equation}
S = \mbox{diag}(e^{i\pi/4},e^{i\pi/4},e^{i\pi/4},e^{i\pi/4},
e^{-i\pi/4},e^{-i\pi/4},e^{-i\pi/4},e^{-i\pi/4})
\end{equation}
Under this symmetry:
\begin{align}
   \Sigma(J_{\bar{n} +}) = SJ_{(-\bar{n}) -}S^{-1}
\end{align}
In particular:
\begin{equation}
\Sigma(J_{\bar{3}+}) = SJ_{\bar{1}-}S^{-1}
\end{equation}
A generic string theory sigma-model does not have any parity symmetry. Parity 
invariance is a property of those backgrounds which only involve the metric, 
axion-dilaton and the RR 5-form field strength, but neither the B-field nor 
the RR 3-form. $AdS_5\times S^5$ is one of such parity-invariant backgrounds.

\subsection{$\beta$-deformations}
The $\beta$-deformations are the simplest deformations of the pure spinor action. 
The corresponding integrated vertex is just the exterior product of two 
global symmetry currents \cite{Mikhailov:2009rx,Bedoya:2010qz}:
\begin{equation}\label{BetaDeformAdSAction}
S_{\rm AdS} \longrightarrow S_{\rm AdS} + \int \varepsilon B^{ab}j_a\wedge j_b
\end{equation}
where $\varepsilon$ is a small parameter measuring the strength of the deformation,
and $B^{ab}$ is a constant super-antisymmetric tensor with indices $a$, $b$
enumerating the generators of the algebra of global symmetries 
${\bf g} = psu(2,2|4)$.
It turns out that when $B$ is of the form $B^{ab} = f^{ab}{}_cA^c$ for some constant $A^c$,
the deformation can be undone by a  field redefinition. Therefore the space 
of linearized $\beta$-deformations is: 
\begin{equation}
{\cal H} = ({\bf g}\wedge {\bf g})/{\bf g}
\end{equation}

\subsection{Physical and unphysical deformations}
Physical $\beta$-deformations have zero internal commutator:
\begin{equation}
{\cal H}_{\rm phys} = ({\bf g}\wedge {\bf g})_0/{\bf g}
\end{equation}
Here $({\bf g}\wedge {\bf g})_0$ means the subspace consisting of $\sum_i \xi_i\wedge \eta_i$ such that:
\begin{equation}\label{PhysicalCondition}
\sum_i [\xi_i, \eta_i] = 0
\end{equation}
Physical deformations describe solutions of linearized SUGRA on the 
background of $AdS_5\times S^5$.

It was explained in \cite{Bedoya:2010qz} that the deformations which belong to the complement 
${\cal H}\backslash {\cal H}_{\rm phys}$ do not correspond to any SUGRA solutions. The spectrum of 
linearized excitations of SUGRA on $AdS_5\times S^5$ does not contain states with 
such quantum numbers.  Attempt to naively identify the supergravity fields 
gives the Ramond-Ramond field strength which is not closed: $dH_{RR} \neq 0$. This
contradicts the SUGRA equations of motion.

For example, consider $B$ of the form:
\begin{equation}\label{BadB}
B^{ab} = 
\left\{
\begin{array}{cl}
f^{ab}_c A^c & \mbox{\small if both $a$ and $b$ are even (bosonic) indices }
\cr
0 & \mbox{\small otherwise}
\end{array}
\right.
\end{equation}
with some constant $A \in so(6) \subset {\bf psu}(2,2|4)$. The corresponding SUGRA
solution would be constant in the AdS directions, and would transform
in the adjoint representation of $so(6)$ (the rotations of the $S^5$).
But there is no such state in the SUGRA spectrum \cite{Kim:1985ez}.

Even without consulting \cite{Kim:1985ez}, that there is no SUGRA solutions with such 
quantum numbers. Let us study the representations of SUGRA fields, even
without equations of motion (off-shell). They are various tensor fields.
A tensor field transforms in some representation $\rho$ of the small algebra
$so(5)\subset so(6)$ (we are looking only at the $S^5$ part). According to the
Frobenius reciprocity, a representation of $so(6)$ enters as many times
as $\rho$ enters into its restriction on $so(5)$. In particular, the adjoint 
representation of $so(6)$ decomposes as follows:
\begin{equation}
\mbox{ad}_{so(6)} = \mbox{ad}_{so(5)} \oplus \mbox{Vec}_{so(5)} 
\end{equation}
But Type IIB SUGRA does not contain vectors, and the only 2-forms are: 
$*_5H_{NSNS}$ and  $*_5H_{RR}$. In the space of 2-forms on $S^5$, the only subspace 
transforming in the adjoint of $so(6)$ are $dX_i\wedge dX_j$ where $S^5$ is parametrized
by $X_1^2+\ldots +X_6^2 = 1$. But $H_{NSNS}$ and $H_{RR}$ are closed 3-forms, while  
$*_5(dX_i\wedge dX_j)$ is not.

\section{Pure spinor formalism in flat space}
\subsection{Action, BRST transformation, supersymmetry and parity}
The action in flat space is:
\begin{align}
S_{\rm flat} = \;
\int d\tau^+ d\tau^- & \;\left[
   {1\over 2}\partial_+x^m\partial_-x^m + p_+\partial_-\theta_L +
   p_-\partial_+\theta_R \;+ 
\right.
\nonumber \\   
&\;\;\; 
\left. 
   +\; w_{+}\partial_-\lambda_L + w_{-}\partial_+\lambda_R
\right]
\label{FlatAction}
\end{align}
where $x,\theta_{L,R}$ are matter fields and $\lambda$ are pure spinor ghosts, and $p_{\pm},w_{\pm}$ 
are their conjugate momenta. The BRST transformation is generated by the
BRST charge:
\begin{equation}\label{DefBRSTCharge}
q_{\rm flat} = \int d\tau^+ \lambda_L d_+ + \int d\tau^- \lambda_R d_-
\end{equation}
where $d_{\pm}$ is some composed field built from $p_{\pm},\;\theta,\;\partial_{\pm}x$, the explicit
expressions are in Section \ref{sec:RelationJAndD}. The corresponding symmetry (called ``BRST 
transformation'')  acts in the following way:
\begin{align}
\epsilon Q_{\rm flat}\; \theta_{L,R} =&\; \epsilon\lambda_{L,R} 
\nonumber\\    
\epsilon Q_{\rm flat}\; x^m = &\; {1\over 2} \left( 
   (\epsilon\lambda_L\Gamma^m\theta_L) + (\epsilon\lambda_R\Gamma^m\theta_R)
\right)
\nonumber\\   
\epsilon Q_{\rm flat}\; \lambda_{L,R} = &\; 0
\nonumber\\   
\epsilon Q_{\rm flat}\; w_{\pm} = &\; \epsilon d_{\pm}
\nonumber \\   
\epsilon Q_{\rm flat}\; d_+ = &\;  \Pi^m_+\Gamma_m\epsilon\lambda_L
\nonumber \\   
\epsilon Q_{\rm flat}\; d_- = &\;  \Pi^m_-\Gamma_m\epsilon\lambda_R
\label{QFlat}
\end{align}
or in compact notations:
\begin{align}
\epsilon Q_{\rm flat} = &\; \epsilon\lambda_L {\partial\over\partial\theta_L}
+ \epsilon\lambda_R {\partial\over\partial\theta_R} 
+ {1\over 2} \left( 
   (\epsilon\lambda_L\Gamma^m\theta_L) + (\epsilon\lambda_R\Gamma^m\theta_R)
\right){\partial\over\partial x^m} \; +
\nonumber \\   
&\; 
+ \epsilon d_{+} {\partial\over\partial w_{+}}
+ \epsilon d_{-} {\partial\over\partial w_{-}}
+ (\Pi^m_+\Gamma_m\epsilon\lambda_L)_{\hat{\alpha}}{\partial\over\partial d_{\hat{\alpha} +}}
+ (\Pi^m_-\Gamma_m\epsilon\lambda_R)_{\alpha}{\partial\over\partial d_{\alpha -}}
\end{align}
\ifodd\amshow 
%  {\tt \href{notes_extra.pdf#href:ActionOfQOnJ}{See notes} } 
\fi
Note that we use the letter $Q$ for both the conserved charge and the
corresponding symmetry action; we hope that this will not lead to confusion.
The BRST operator $\epsilon Q_{\rm flat}$ has the following key properties:
\begin{enumerate}
\item It is a symmetry of the action
\item It is nilpotent: $Q_{\rm flat}^2 = 0$ (up to gauge tranformations)
\end{enumerate}
Besides the BRST invariance, the flat space action is also invariant under
the super-Poincare transformations. In particular, there are supersymmetries
$t^3_{\alpha}$ and $t^1_{\dot{\alpha}}$ which act as follows:
\begin{align}
\kappa_L^{\alpha} t_{\alpha}^3 = &\; \kappa_L^{\alpha}{\partial\over\partial \theta_L^{\alpha}} - {1\over 2}
   (\kappa_L\Gamma^m\theta_L) {\partial\over\partial x^m}
\nonumber \\    
\kappa_R^{\hat{\alpha}} t_{\hat{\alpha}}^1 = &\; \kappa_R^{\hat{\alpha}}{\partial\over\partial \theta_R^{\hat{\alpha}}} - {1\over 2}
   (\kappa_R\Gamma^m\theta_R) {\partial\over\partial x^m}
\label{SUSY}
\end{align}
where $\kappa_L^{\alpha}$ and $\kappa_R^{\hat{\alpha}}$ are constant Grassmann numbers, enumerating the SUSY 
generators. 

The flat space theory has parity invariance, as Eq. (\ref{AdSParity}) of $AdS_5\times S^5$.
It exchanges $\tau^+$ with $\tau^-$ and $\theta_L$ with $\theta_R$. 

\subsection{Using AdS notations in flat space}\label{sec:AdSNotations}
Even in the strict flat space limit, it is still convenient to use the AdS 
notations. For example:
\begin{align}
[\theta_L,\partial_+\theta_L]^m =&\; (\theta_L\Gamma^n\partial_+\theta_L)
\\     
[\theta_L,\theta_R]^{[mn]} = &\; (\theta_L F^{mnpqr}\Gamma_{pqr} \theta_R)
\\  
[B_2, \theta_L]^{\hat{\alpha}} = &\; 
\left(\widehat{F} B_2^m\Gamma_m \theta_L\right)^{\hat{\alpha}}
\end{align}
where $F^{mnpqr}$ is the RR 5-form field strength of $AdS_5\times S^5$ in the flat space
limit. We will also put ${\bf Z}_4$ indices on the currents; the Lorentz currents will
be denoted $j_{0\pm}$, the translations $j_{2\pm}$, and the supersymmetries $j_{3\pm}$ and $j_{1\pm}$.

\section{Deformations of the flat space structures}
\subsection{Deforming $Q_{\rm flat}$ keeping $S_{\rm flat}$ undeformed}

\subsubsection{Construction of the deformation}
\label{sec:DeformQ}
Consider the following infinitesimal deformation of the BRST charge,
parametrized by the constant bispinors $B^{\hat{\alpha}\hat{\beta}}_R$ and $B_L^{\alpha\beta}$:
\begin{align}
   \epsilon q_B = &\; \epsilon q_{\rm flat} + \epsilon \Delta_{\rm wild}q 
\nonumber \\  
\mbox{\tt\small where }\epsilon \Delta_{\rm wild}q
= &\; \epsilon q_{\rm flat}  + \varepsilon\int 
\left(
   (\theta_L\Gamma_m\epsilon\lambda_L)
   \Gamma^m_{\alpha\gamma}\theta_L^{\gamma}
\right)B_L^{\alpha\beta} S_{\beta+}d\tau^+ \;+
\nonumber \\     
&\;\phantom{\epsilon q_{\rm flat}} + \varepsilon\int 
\left(
   (\theta_R\Gamma_m\epsilon\lambda_R)
   \Gamma^m_{\hat{\alpha}\hat{\gamma}}\theta_R^{\hat{\gamma}}
\right)B_R^{\hat{\alpha}\hat{\beta}} S_{\hat{\beta}-}d\tau^- 
\end{align}
Notations:
\begin{itemize}
\item  $q_{\rm flat}$ is the standard flat space BRST charge (\ref{DefBRSTCharge}).
\item  $B^{\hat{\alpha}\hat{\beta}}_R$ and $B_L^{\alpha\beta}$ are constant bispinors, $B_L^{\alpha\beta} = B_L^{\beta\alpha}$, $B_R^{\hat{\alpha}\hat{\beta}} = B_R^{\hat{\beta}\hat{\alpha}}$.
\item  $\varepsilon$ is a small parameter, measuring the strength of the deformation; it 
   should not be confused with $\epsilon$ --- the formal Grassmann number. Note that 
   $\varepsilon$ is bosonic and $\epsilon$ is fermionic. To the first order in $\varepsilon$ the 
   deformed BRST operator is a new nilpotent symmetry of the action. 
\item  $S_{\beta +}$ and $S_{\hat{\beta}-}$ are the holomorphic (left) and the antiholomorphic (right) 
   supersymmetry charges\footnote{The fact that the
     supersymmetry charges are holomorphic or antiholomorphic is special to
     flat space, and is crucial for our construction} (see Eqs. (\ref{S1Plus}) and (\ref{S3Minus}) for the explicit formulas)
\end{itemize}
It follows from the definition that $\Delta_{\rm wild}q$ is a conserved charge. Indeed,
on-shell $\partial_-S_{\beta+} = \partial_+S_{\hat{\beta}-} = 0$ and $\partial_-\theta_L = \partial_+\theta_R = \partial_-\lambda_L = \partial_+\lambda_R = 0$. 

The deformation $\Delta_{\rm wild}q$ consists of the ``left'' piece (proportional to $B_L$)
and the ``right'' piece (proportional to $B_R$). These two pieces provide two
separate deformations, the left one and the right one. They are separately
well-defined.

\subsubsection{Proof that $\Delta_{\rm wild} q$ anticommutes with $q_{\rm flat}$}
We will prove this using the Hamiltonian formalism. Let us calculate the
Poisson bracket:
\begin{equation}
   \{ q_{\rm flat},\;\Delta_{\rm wild}q \} = Q_{\rm flat}\;\Delta_{\rm wild}q
\end{equation}
Notice the descent relation for the density of $\Delta_{\rm wild}q$:
\begin{align}
&\; \epsilon Q_{\rm flat}  
\left(\;\left(
   (\theta_L\Gamma_m\epsilon'\lambda_L)
   \Gamma^m_{\alpha\gamma}\theta_L^{\gamma}
\right)B_L^{\alpha\beta} S_{\beta+}\;\right) =
\nonumber  \\   
= &\; \partial_+ \left( {1\over 6} \left(
   (\theta_L\Gamma_m\epsilon'\lambda_L)
   \Gamma^m_{\alpha\gamma}\theta_L^{\gamma}
\right)B_L^{\alpha\beta} 
\left(
   (\theta_L\Gamma_m\epsilon\lambda_L)
   \Gamma^m_{\beta\delta}\theta_L^{\delta}
\right)
\right)   
\end{align}
which follows from the descent of the SUSY current:
\begin{align}
&\; \epsilon Q_{\rm flat}  S_{\alpha +} =
\partial_+ \left( {1\over 3} 
   (\theta_L\Gamma_m\epsilon\lambda_L)
   \Gamma^m_{\alpha\gamma}\theta_L^{\gamma}
\right)
\end{align}
which can be derived by an explicit calculation, or as a limit of the similar
relation in the $AdS_5\times S^5$ sigma-model derived in \cite{Berkovits:2004jw}  and reviewed in \cite{Bedoya:2010qz}.
\ifodd\amshow
%  \href{notes_extra.pdf#DescentOfTLTLLL}{Explanation of $1\over 6$}
\fi
Let us introduce the notation:
\begin{equation}
v_{L\alpha}(\epsilon) = \left(
   (\theta_L\Gamma_m\epsilon\lambda_L)
   \Gamma^m_{\alpha\gamma}\theta_L^{\gamma}
\right)
\end{equation}
With this notations we have:
\begin{equation}\label{DescentOfDeformedChargeDensity}
\epsilon Q_{\rm flat}\left(
   v_{L\alpha}(\epsilon')B_L^{\alpha\beta}S_{\beta+}
\right) = {1\over 6}\partial_+\left(
   v_{L\alpha}(\epsilon')B_L^{\alpha\beta}v_{L\beta}(\epsilon)
\right)
\end{equation}
There is a similar descent relation for the charge density of the right 
deformation. Eq. (\ref{DescentOfDeformedChargeDensity}) means that the $Q_{\rm flat}$-variation of the density of
$\Delta_{\rm wild}q$ is a total derivative, and this implies:
\begin{equation}
   \{ q_{\rm flat},\;\Delta_{\rm wild}q \} = Q_{\rm flat}\;\Delta_{\rm wild}q = 0
\end{equation}

\subsubsection{Deformation of the BRST transformation}
This deformation of the BRST charges corresponds to the following deformation
of the BRST transformation:
\begin{align}\label{DeformQ}
\epsilon Q_B =\;& \epsilon Q_{\rm flat} + \Delta_{\rm wild}Q
\\   
\mbox{\tt\small where }\Delta_{\rm wild}Q = \;&
\varepsilon B^{\hat{\alpha}\hat{\beta}}_R
\left(
   (\theta_R\Gamma_m\epsilon\lambda_R)
   \Gamma^m_{\hat{\alpha}\hat{\gamma}}\theta_R^{\hat{\gamma}} 
\right)\; t^1_{\hat{\beta}}
+ \varepsilon B^{\alpha\beta}_L
\left(
   (\theta_L\Gamma_m\epsilon\lambda_L)
   \Gamma^m_{\alpha\gamma}\theta_L^{\gamma}
\right)\; t^3_{\beta} \;+ 
\nonumber
\\   
&\; + k_{\alpha+}{\partial\over\partial p_{\alpha+}}
+ l_{\alpha+}{\partial\over\partial w_{\alpha +}}
+ k_{\hat{\alpha}-}{\partial\over\partial p_{\hat{\alpha}-}}
+ l_{\hat{\alpha}-}{\partial\over\partial w_{\hat{\alpha}-}}
\end{align}
where $t^1_{\hat{\beta}}$ and $t^3_{\beta}$ are the right and left SUSY generators given by Eq. (\ref{SUSY}), 
and $k_{\alpha+},l_{\alpha+},k_{\hat{\alpha}-},l_{\hat{\alpha}-}$ define some infinitesimal shifts of the 
momenta $p_{\pm}, w_{\pm}$. We will not need the explicit formula for these shifts; 
they are canonically defined in terms of the shifts of $x$ and $\theta$ generated
by $t^1_{\hat{\beta}}$ and $t^3_{\beta}$.

\subsubsection{When such a deformation can be undone by a field redefinition?}
\label{sec:SpecialCaseB}
\paragraph     {Sufficient condition}
Consider the special case when $B_L$ satisfies:
\begin{equation}\label{IntCommZeroLeft}
\Gamma^m_{\alpha\beta}\; B_L^{\alpha\beta} = 0
\end{equation}
In this case exists $W_L$:
\begin{equation}\label{ExistsWL}
   v_{L\alpha}(\epsilon')B_L^{\alpha\beta}v_{L\beta}(\epsilon) =
\epsilon Q_{\rm flat}\left(\epsilon' W_L\right)
\end{equation}
The structure of $W_L$ is $[\theta^5_L\lambda_L]$. This implies:
\begin{equation}
Q_{\rm flat}\left(
   v_{L\alpha}(\epsilon')B_L^{\alpha\beta} S_{\beta+} - \partial_+(\epsilon' W_L)
\right) = 0
\end{equation}
Because the cohomology in conformal dimension $1$ is trivial, this implies the 
existence of $y_{L+}$:
\begin{equation}
   v_{L\alpha}(\epsilon')B_L^{\alpha\beta} S_{\beta+} =
\partial_+(\epsilon' W_L) + \epsilon' Q_{\rm flat}y_{L+}
\end{equation}
(See the discussion in Appendix \ref{sec:AppendixConfDimOne}.)
\ifodd\amshow
%  {\tt \href{notes_extra.pdf#href:GenCommentsOnDescent}{on descent}}
%  {\tt \href{notes_extra.pdf#href:CommentOnConfDim1}{on conf. dim. 1}}
\fi
We observe that $\partial_-y_{L+} \onshellflat 0$. Thus $y_{L+}$ is a conserved  current of the flat 
space theory generating some transformation $Y_L$. We have therefore:
\begin{equation}\label{UndoneByFieldRedefinition}
Q_B = Q_{\rm flat} + [Y_L,Q_{\rm flat}]
\end{equation}
Therefore if (\ref{IntCommZeroLeft}) then the deformation $Q_B\to Q_{\rm flat}$ is trivial. 

\paragraph     {Necessary condition}
Let us assume that exists a  vector field $Y_L$ satisfying Eq. (\ref{UndoneByFieldRedefinition}). Let us 
assume that $Y_L$ is a symmetry of the $S_{\rm flat}$; in the next Section \ref{sec:AltProofOfNecessaryCondition} we will 
give a proof without this assumption. 
\ifodd\amshow
%\href{notes_extra.pdf#href:TradeForVertex}{Comment}
\fi
Then the conserved current $v_{L\alpha}B_L^{\alpha\beta} S_{\beta+}$ corresponding to $Q_B^{(1)}$ satisfies:
\begin{equation}
 v_{L\alpha}B_L^{\alpha\beta} S_{\beta+} = Y_L j_{\rm flat\; BRST+} + \partial_+ \phi
\end{equation}
for some holomorphic $\phi$. Using that $Q_{\rm flat}j_{\rm flat\; BRST+} = 0$, this implies:
\begin{align}
   Q_{\rm flat}\left( v_{L\alpha}B_L^{\alpha\beta} S_{\beta+} \right) = 
   \partial_+\left(Q_{\rm flat}\phi\right)
\end{align}
Therefore:
\begin{equation}
v_{L\alpha}B_L^{\alpha\beta} v_{L\beta} = Q_{\rm flat}\phi
\end{equation}
In the rest of this paragraph we will prove that this is only possible 
when (\ref{IntCommZeroLeft}). Indeed, suppose that (\ref{IntCommZeroLeft}) is not satisfied.
Without loss of generality, we can assume: $B_L^{\alpha\beta} = B^m\Gamma_m^{\alpha\beta}$. We want to prove
that $(v_{L}\hat{B} v_{L})$ represents a nonzero cohomology class of $Q_{\rm flat}$. Remember that
$Q_{\rm flat}$ is defined in (\ref{QFlat}). Let us formally split $x$ into $x_L$ and $x_R$:
\begin{align}
   x^m = x_L^m + x_R^m
\\   
   \epsilon Q_{\rm flat} x_L^m = {1\over 2}(\epsilon\lambda_L\Gamma^m\theta_L)
\\    
\epsilon Q_{\rm flat} x_R^m = {1\over 2}(\epsilon\lambda_R\Gamma^m\theta_R)
\end{align}
Let us extend the BRST complex\footnote{I want to thank M.~Movshev for teaching me this trick} by including functions of $x_L$ and $x_R$ 
(and not just of their sum). Then $(v_{L}\hat{B} v_{L})$ is BRST trivial:
\begin{align}\label{DefCalA}
   (v_{L}\hat{B} v_{L}) =\;& Q_{\rm flat}{\cal A}
\\    
\mbox{ \small where\; }{\cal A}= \;&
      A_m(x_L)(\theta_L\Gamma^m\lambda_L) + (dA)_{mn}[\theta_L^3\lambda_L]^{[mn]}
      +\ldots 
\end{align}
where $A_m(x_L)$ is such that:
\begin{equation}\label{ConstantChargeDensity}
   d*dA = *B
\end{equation}
In other words, $A$ is the Maxwell field created by the constant charge
density $B$. The question is:
\begin{itemize}
\item Is it possible to correct ${\cal A}$ by adding to it something $Q_{\rm flat}$-closed, so 
that the corrected ${\cal A}$ depends on $x_L$ and $x_R$ only through $x=x_L+x_R$?
\end{itemize}
If this is possible then  $(v_{L}\hat{B} v_{L})$ is $Q_{\rm flat}$ exact. 
We will now prove that it is not possible to make such a correction of ${\cal A}$,
and therefore  $(v_{L}\hat{B} v_{L})$ is cohomologically nontrivial. 

\hangindent3ex
\hangafter=0
\noindent
A function of $x_L,x_R,\theta_L,\theta_R,\lambda_L,\lambda_R$ can be written in terms of 
$x,\theta_L,\theta_R,\lambda_L,\lambda_R$ if and only if it is annihilated by $y^m \left( 
   {\partial\over\partial x_L^m} - {\partial\over\partial x_R^m}
\right)$ for any
constant vector $y^m$. Notice that:
\begin{align}
 Q_{\rm flat}\left[y^m \left( 
   {\partial\over\partial x_L^m} - {\partial\over\partial x_R^m}
\right)\;{\cal A}\;\right] = 0
\label{DiffXLXR}
\end{align}
--- this is because $y^m\left( 
   {\partial\over\partial x_L^m} - {\partial\over\partial x_R^m}
\right)$ commutes with $Q_{\rm flat}$ and annihilates 
$(v_L\hat{B}v_L)$. Let us consider the following solution of (\ref{ConstantChargeDensity}):
\begin{equation}
   A_{\mu} = {1\over 18} x^2 B_{\mu}
\end{equation}
Then $F_{\mu\nu} = {1\over 9}\left(x_{\mu}B_{\nu} - x_{\nu} B_{\mu}\right)$. We see that $\left(y\partial_{x_L} - y\partial_{x_R}\right){\cal A}$ represents a 
nontrivial cohomology class of $Q_{\rm flat}$, corresponding to the Maxwell field of 
the constant field strength $y\wedge B$. Now the question is:
\begin{itemize}
\item Is it possible to obtain this cohomology class by acting with 
   $(y\partial_{x_L} - y\partial_{x_R})$ on some cohomology class ${\cal Z}$ of $Q_{\rm flat}$?
\end{itemize}
In other words, is it possible that exists ${\cal Z}$ such that:
\begin{align}\label{AandAPrime}
   (y\partial_{x_L} - y\partial_{x_R}){\cal A} = \;& 
   (y\partial_{x_L} - y\partial_{x_R}) {\cal Z}
\\    
   Q_{\rm flat} {\cal Z} =\;& 0
\label{ZIsClosed}
\end{align}
(such a ${\cal Z}$ will necessarily be nontrivial in the cohomology of $Q_{\rm flat}$)?
If and only if this were possible, then we could modify ${\cal A}$ by subtracting 
from it a representative of ${\cal Z}$ (and since $\cal Z$ is closed, this will not change 
the defining property (\ref{DefCalA})) so that the modified ${\cal A}$ depends on $x_L$ and $x_R$ 
through $x=x_L+x_R$. Then Eq. (\ref{DefCalA}) would have implied that $(v_L\hat{B}v_L)$ is
BRST exact. We will now prove that this is impossible. 

\hangindent6ex
\hangafter=0
\noindent
Suppose that exists ${\cal Z}$ such that (\ref{AandAPrime}) and (\ref{ZIsClosed}). As we already said, since
the $Q_{\rm flat}$-cohomology class of  $\left(y\partial_{x_L} - y\partial_{x_R}\right){\cal A}$ is nontrivial, $\cal Z$ should be 
also nontrivial in $Q_{\rm flat}$-cohomology. Modulo $Q_{\rm flat}$-exact terms $\cal Z$ has to be of 
the following form:
\begin{align}\label{CalAPrime}
   {\cal Z} = Z_{Lm}(x_L,x_R) (\theta_L\Gamma^m\lambda_L) 
+ Z_{Rm}(x_L,x_R) (\theta_R\Gamma^m\lambda_R) + [x\lambda\theta^3] + [\lambda\theta^5]
\end{align}
where $Z_{Lm}$ and $Z_{Rm}$ are quadratic in $x$. For (\ref{CalAPrime}) to be $Q_{\rm flat}$-closed we 
need:
\begin{align}\label{RelationBetweenDerivativesOfZ}
   \partial_{x_R^n} Z_{Lm} = \partial_{x_L^m} Z_{Rn}
\end{align}
Since both $Z_{Lm}$ and $Z_{Rn}$ are quadratic polynomials in $(x_L,x_R)$, let us 
introduce the notations:
\begin{align}
   Z_{Lm} =\;& Z_{Lm,LL} + Z_{Lm,LR} + Z_{Lm,RR}
\nonumber\\    
Z_{Rm} =\;& Z_{Rm,LL} + Z_{Rm,LR} + Z_{Rm,RR}
\end{align}
where {\it e.g.} $Z_{Rn,LL}$ is the term with $x_L x_L$ in $Z_{Rm}$, {\it etc.}.
Eq. (\ref{RelationBetweenDerivativesOfZ}) implies that the term with $x_Rx_L$ in $Z_{Lm}$  and the term with $x_Lx_L$ 
in $Z_{Rm}$ can be gauged away by $Q_{\rm flat}(2Z_{Rn,LL}x_R^n)$: 
\[
   Z_{Lm,LR}(x_L,x_R) (\theta_L\Gamma^m\lambda_L) 
+ Z_{Rm,LL}(x_L,x_R) (\theta_R\Gamma^m\lambda_R) \;=
Q_{\rm flat}(2Z_{Rn,LL}x_R^n)
\]
Similarly, the terms with $x_Rx_R$ in $Z_{Lm}$ plus terms with $x_Lx_R$ in $Z_{Rm}$ are 
$Q_{\rm flat}(2Z_{Ln,RR} x^n_L)$, where $Z_{Ln,RR}$ is the coefficient of $x_Rx_R$ in $Z_{Ln}$. After 
such a gauge transformation, we are left with:
\begin{align}\label{CalAPrimeGauged}
   {\cal A}' = Z_{Lm}(x_L) (\theta_L\Gamma^m\lambda_L) 
+ Z_{Rm}(x_R) (\theta_R\Gamma^m\lambda_R) + [x\lambda\theta^3] + [\lambda\theta^5]
\end{align}
Now we observe that this corresponds to a pair of Maxwell fields with the
field strength linearly dependent on the spacetime coordinates. One of these 
two Maxwell fields corresponds to $Z_{Lm}$, and another to $Z_{Rm}$. Up to gauge
transformations, both transform in the traceless part of the 
\begin{picture}(21,21)(0,4)
\put(0,0){\framebox(10,10){}}
\put(10,10){\framebox(10,10){}}
\put(0,10){\framebox(10,10){}}
\end{picture} of $so(1,9)$. At the same time, the cohomology class of 
$(y\partial_{x_L}-y\partial_{x_R}){\cal A}$ is parametrized by the vector $B$, therefore it transforms in
a vector ({\it i.e.} \begin{picture}(11,11)(0,0)
\put(0,0){\framebox(10,10){}}
\end{picture}) of $so(1,9)$. This implies that (\ref{AandAPrime}) is impossible.

\subsubsection{Another proof of the necessary condition for triviality}
\label{sec:AltProofOfNecessaryCondition}
Let us take $B^{\alpha\beta} = B^m\Gamma_m^{\alpha\beta}$. Suppose that exists an infinitesimal field 
redefinition $Y_L$ such that (\ref{UndoneByFieldRedefinition}). Let us study the action of $Y_L$ on $\lambda_L$. 
We observe: 
\begin{align}
Q_{\rm flat}\theta_L = \;& \lambda_L
\\   
   (Q_B-Q_{\rm flat})\theta_L = \;&
   \hat{B}\Gamma^m\theta_L(\theta_L\Gamma^m\lambda_L)
\end{align}
Therefore in order to satisfy (\ref{UndoneByFieldRedefinition}) we should have:
\begin{align}\label{PossibleYLLambda}
   Y_L\lambda_L = \hat{B}\Gamma^m\theta_L(\theta_L\Gamma^m\lambda_L) + 
   Q_{\rm flat}\Xi
\end{align}
for some $\Xi$ (we have $\Xi=Y_L\theta_L^{\alpha}$). Moreover, we should satisfy the pure spinor 
constraint:
\begin{align}\label{PureSpinorConstraintVariation}
   (\lambda_L\Gamma^k  Y_L\lambda_L) = 0
\end{align}
Notice that $Y_L\lambda_L$ is necessarily $Q_{\rm flat}$-closed, and that $\Xi$ is necessarily of 
the form $[\theta^3B]$. The only expression of the form $[\theta^2\lambda B]$ which satisfies (\ref{PureSpinorConstraintVariation})
would have been:
\begin{equation}
   Y_L\lambda_L = \Gamma^{mn}\lambda B^l(\theta\Gamma_{lmn}\theta) 
\end{equation}
but this is not BRST closed and therefore is not of the form (\ref{PossibleYLLambda}).
\ifodd\amshow
%{\tt \href{notes_extra.pdf#href:GammaMNLambdaThetaGammaLMNTheta}{Calculation}}
\fi

\paragraph     {Comment} 
But when $B$ is a 5-form rather than a vector, see Eq. (\ref{YLFor5Form}).

\paragraph     {Conclusion}
Eq. (\ref{IntCommZeroLeft}) is a necessary and sufficient condition for the 
triviality of the deformation. In other words, the deformation of the flat 
space BRST operator parametrized by $B_L^{\alpha\beta}$ can be undone by a 
symmetry of the  action if and only if (\ref{IntCommZeroLeft}).

\subsubsection{Extension to higher orders}
\label{sec:ExtensionToHigherOrders}
It should be possible to extend the deformation (\ref{DeformQ}) to higher orders in $\varepsilon$.
Let us for now put $B_R = 0$ in (\ref{DeformQ}); that is, restrict ourselves to the 
``left'' deformations only. We get:
\begin{align}
   \{Q_B,Q_B\} =\;& \varepsilon^2 
   \left( 
      (\theta\Gamma^m\lambda)
      \left(
         \theta\Gamma^m B{\partial\over\partial\theta}
      \right)
   \right)^2 \;=
   \nonumber \\    
   =\;& \varepsilon^2 
   (\theta\Gamma^m\lambda)(\theta\Gamma^m B\Gamma^n\lambda)
      \left(
         \theta\Gamma^n B{\partial\over\partial\theta}
      \right)\; -
   \nonumber \\
   & - \varepsilon^2
   (\theta\Gamma^m\lambda)(\theta\Gamma^n\lambda)
   \left(
      \theta\Gamma^m B \Gamma^n B {\partial\over\partial\theta}
   \right)
\label{QBQB}
\end{align}
If $B$ is a 5-form, then one can see that this is BRST exact; but in fact we 
have already seen in Section \ref{sec:SpecialCaseB} that in this case $Q_B$ is a trivial 
deformation of $Q_{\rm flat}$. If $B$ is a 1-form, then the obstacle is proportional 
to $B_mB_m$. To calculate the coefficient, we observe: 
\ifodd\amshow
for any odd $\eta$:
\begin{equation}
   (\eta\Gamma^n \epsilon\lambda)\theta \Gamma^n +
   (\epsilon\lambda\Gamma^n \theta)\eta \Gamma^n +
   (\theta\Gamma^n \eta)\epsilon\lambda \Gamma^n = 0
\end{equation}
Therefore up to a BRST-exact expression:
\begin{equation}
   (\eta\Gamma^n \epsilon\lambda)\theta \Gamma^n = 
   {1\over 2}(\theta\Gamma^n\epsilon\lambda)\eta\Gamma^n + Q_{\rm flat}(\ldots)
\end{equation}
This allows us to transform:
\fi
\begin{align}
   (\theta\Gamma^m\lambda)(\theta\Gamma^m B\Gamma^n\lambda)\theta\Gamma^n 
   = {1\over 2}
   (\theta\Gamma^m\lambda)(\theta\Gamma^n \lambda)\theta\Gamma^m B \Gamma^n
+ Q_{\rm flat}(\ldots)
\end{align}
This means:
\begin{equation}
   \{Q_B,Q_B\} = - {1\over 2} \varepsilon^2
   (\theta\Gamma^m\lambda)(\theta\Gamma^n\lambda)
   \left(
      \theta\Gamma^m B \Gamma^n B {\partial\over\partial\theta}
   \right) + [Q_{\rm flat},\ldots]
\end{equation}
In $Q_{\rm flat}$ cohomology this is proportional to $B_mB_m$. To calculate the 
coefficient of proportionality we can substitute $B\otimes B = \Gamma^k\otimes \Gamma^k$. We get:
\begin{equation}
   \{Q_B,Q_B\} =  {2\over 5} |B|^2\varepsilon^2
   (\theta\Gamma^m\lambda)(\theta\Gamma^n\lambda)
   \left(
      \theta\Gamma^{mn}  {\partial\over\partial\theta}
   \right) + [Q_{\rm flat},\ldots]
\end{equation}
Where $|B|^2 = B_mB_m$. When $B$ is a lightlike vector, we can construct $Q_B^{(2)}$ 
such that the operator:
\begin{align}
Q'_B = Q_{\rm flat} + 
\varepsilon (\theta\Gamma^m\lambda)
      \left(
         \theta\Gamma^m B{\partial\over\partial\theta}
      \right)\;+\; \varepsilon^2 Q_B^{(2)}
\end{align}
which is nilpotent up to the terms of the order $\varepsilon^3$. One can continue this
procedure to higher orders in $\varepsilon$. The only invariant which can arise is $|B|^2$.
Therefore we conclude that the deformation $Q_{\rm flat} \to Q_B$ is unobstructed
when $B$ is lightlike, {\it i.e.} $|B|^2 =0$.

\subsubsection{Relation to $\beta$-deformation}
\label{sec:RelationToBeta}
The deformation of the AdS action given by (\ref{BetaDeformAdSAction}) preserves the BRST
invariance of the action, but actually changes the action of the BRST
transformation. Indeed, the deforming vertex is only BRST-closed on-shell:
\begin{equation}\label{AdSDeformationQClosedOnShell}
Q_{\rm AdS} \left(\int B^{ab}j_a\wedge j_b\right) \onshellads 0
\end{equation}
where $\onshellads$ means ``up to the equations of motion of the AdS $\sigma$-model''.
Because (\ref{AdSDeformationQClosedOnShell}) only holds on-shell, the
deformed action is not invariant under the orginal BRST transformation,
but instead under a deformed  BRST transformation. The necessary deformation 
of the BRST transformation was constructed in \cite{Bedoya:2010qz}, where it was called $Q_1$:
\begin{equation}\label{Q1}
\epsilon Q_1 = 
4\left(g^{-1}(\epsilon\lambda_3  - \epsilon\lambda_1)g\right)_a \;B^{ab}\;t_b
\end{equation}
Here $t_b$ are generators of ${\bf g} = {\bf psu}(2,2|4)$.  Expanding $\left(g^{-1}(\epsilon\lambda_3  - \epsilon\lambda_1)g\right)_{\bar{1}}$ in 
powers of $x$ and $\theta$, we get:
\begin{align}\label{QandPsi}
 \left(g^{-1}(\epsilon\lambda_3  - \epsilon\lambda_1)g\right)_{\bar{1}} =
\;& 
\epsilon Q \Psi_1 
- {4\over 3}[\theta_L,[\theta_L,\epsilon\lambda_L]] \; + \;\ldots
\\[5pt] 
\mbox{\tt \small where } \Psi_1 = \;&
      - \theta_R - [x,\theta_L] + {1\over 3}[\theta_R,[\theta_R,\theta_L] ]
\end{align}
where dots stand for the higher order terms. Similarly:
\begin{align}\label{QandPsi3}
 \left(g^{-1}(\epsilon\lambda_3  - \epsilon\lambda_1)g\right)_{\bar{3}} =
\;& 
\epsilon Q \Psi_3
+ {4\over 3}[\theta_R,[\theta_R,\epsilon\lambda_R]] \; + \;\ldots
\\[5pt] 
\mbox{\tt \small where } \Psi_3 = \;&
      \theta_L + [x,\theta_R] - {1\over 3}[\theta_L,[\theta_L,\theta_R] ]
\end{align}
We conclude that:
\begin{itemize}
\item up to a BRST exact expression  $\epsilon Q_1$ is identical to $\epsilon Q_B$ of (\ref{DeformQ}).
This means that the leading effect in the flat space limit of this particular 
nonphysical $\beta$-deformation is to deform the BRST structure of the flat space 
action as in Eq \eqref{DeformQ}.
\end{itemize}     

\subsubsection{Field reparametrization $K$}
\label{sec:FieldReparametrizationK}
Let us consider a particular example of $B^{ab}$, when the only nonzero 
component has both upper indices $a$ and $b$ in ${\bf g}_1$, and $B$ has the form:
\begin{equation}
B^{\hat{\alpha}\hat{\beta}} = f^{\hat{\alpha}\hat{\beta}}{}_m B_{L2}^m
\end{equation}
In this case:
\begin{align}
   \epsilon Q_1 =\;& 
-{16\over 3}[\;B_{L2}\;,\;[\theta_L,[\theta_L,\epsilon\lambda_L]]\;]^{\alpha}
\; t^3_{\alpha}  + [\epsilon Q\;,\;K_L]
\\     
\mbox{\tt \small where } K_L=\;& 4[B_{L2},\Psi_1]^{\alpha}\;t^3_{\alpha}
\label{DefKLeft}
\end{align}
This means that $\epsilon Q_1$ is of the form (\ref{DeformQ}) after a field reparametrization 
specified by the vector field $K_L$.  

Similarly, consider the case when the only nonzero components of $B^{ab}$ are the 
following: 
\begin{equation}
B^{\alpha\beta} = f^{\alpha\beta}{}_m B_{R2}^m
\end{equation}
In this case:
\begin{align}
   \epsilon Q_1 =\;& 
{16\over 3}[\;B_{R2}\;,\;[\theta_R,[\theta_R,\epsilon\lambda_R]]\;]^{\hat{\alpha}}
\; t^1_{\hat{\alpha}}  + [\epsilon Q\;,\;K_R]
\\     
\mbox{\tt \small where } K_R=\;& 4[B_{R2},\Psi_3]^{\hat{\alpha}}\;t^1_{\hat{\alpha}}
\label{DefKRight}
\end{align}

\paragraph     {Action of $K$ on $S_{\rm AdS}$}
\begin{align}
   KS_{\rm AdS} =\;& 
   -\int d^2\tau \;\mbox{Str}\left(
      \partial_+[B_2,\Psi_1] \; j_{1-} - \partial_-[B_2,\Psi_1] \; j_{1+}
   \right)
\label{KSAdS}
\end{align}
{\small Observe that  $j_{1-} = -\partial_-\theta_R+\ldots$ and $j_{1+} = 3\;\partial_+\theta_R +\ldots$
With our definition of $j_{\pm}$ we have:
\begin{equation}
\xi . S_{\rm AdS} = - {1\over 4}\int d^2\tau \; \mbox{Str} \left(
   \partial_+ \xi\; j_- - \partial_- \xi\; j_+
\right)
\end{equation}
}

\subsection{Deforming ($S_{\rm flat}$, $Q_{\rm flat}$) to ($S_{\rm AdS}$, $Q_{\rm AdS}$)}
Going from flat space to AdS changes the action, by turning on the RR
five-form field strength. To describe the corresponding deformation of the 
action it is useful to introduce a small parameter $1/R$, which corresponds
to the inverse radius of the AdS space. The scaling of the basic fields
is as follows:
\begin{align}
&   x \simeq R^{-1},\quad
   \theta_{L,R} \simeq R^{-1/2},\quad
   p_{\pm} \simeq d_{\pm} \simeq R^{-3/2},\quad
\nonumber \\
&   \lambda_{L,R} \simeq R^{-1/2},\quad
   w_{\pm} \simeq R^{-3/2}
\end{align}
With these notations the flat action (\ref{FlatAction}) is of the order $R^{-2}$. (Usually 
there is an overall coefficient $R^{-2}$ in front of the action, then the action 
is of the order $1$. But we will prefer to omit this overall coefficient.)

The RR five-form deforms the action as follows:
\begin{equation}
S_{\rm flat}\to S_{\rm flat} + \int F^{\alpha\hat{\beta}}d_{\alpha}d_{\hat{\beta}} + \ldots
\end{equation}
where $\ldots$ is for terms containing $\theta$. We observe that the deformation term 
is of the order $R^{-3}$ (while the $S_{\rm flat}$ is of the order $R^{-2}$). 

We will denote the AdS deformation vertex $U_{AdS}$:
\begin{equation}\label{UAdSviaD}
U_{\rm AdS} = F^{\alpha\hat{\beta}}d_{\alpha}d_{\hat{\beta}} + \mbox{ \small [terms with $\theta$] }
\end{equation}
(The complete formula is (\ref{CompleteURR5Form}).) Once again, observe that the flat space 
action is of the order $R^{-2}$, and the deformation $U_{\rm AdS}$ is of the order $R^{-3}$.

\subsection{Interplay between the two deformations}
We have considered two deformations of the flat space superstring: the 
deformation (\ref{DeformQ}) which leaves the action invariant and only changes the BRST 
structure, and the deformation from flat space to $AdS_5\times S^5$. Let us look at 
the interplay between these two deformations.
The action of $Q_{\rm flat}$ on $U_{AdS}$ is a total derivative on the equations of motion 
of $S_{\rm flat}$. But the deformed $Q$  generally speaking acts nontrivially:
\begin{align}
Q_{\rm flat}\int U_{AdS} & \onshellflat 0
\\   
Q_B\int U_{AdS} & \onshellflat  R^{-4}
\label{QBOnURR5FormSchematic}
\end{align}
where $ \onshellflat $ means equality up to the equations of motion of flat space. 
In the next Section we will see that (\ref{QBOnURR5FormSchematic}) is important for understanding the
flat space limit. 

\section{Flat space limit of the $\beta$-deformation vertices}

\subsection{Flat space limit of the $AdS_5\times S^5$ sigma-model}
\subsubsection{Coset space and BRST operator}
We choose the following parametrization of the 
$PSU(2,2|4)/(SO(1,4)\times SO(5))$ coset space:
\begin{equation}
g = e^{\theta} e^X
\end{equation}
The action of the  BRST operator on the matter fields:
\begin{equation}
   \epsilon Q g \; = \; \epsilon(\lambda_L + \lambda_R)g + \omega(\epsilon)g
\end{equation}
where $\omega(\epsilon)$ is some compensating $SO(1,4)\times SO(5)$ gauge transformation.

\vspace{10pt}

{\small \noindent In terms of $\theta$ and $x$:
\begin{align}
\epsilon Q =&\; \epsilon\lambda_L {\partial\over\partial \theta_L} +
    \epsilon\lambda_R {\partial\over\partial \theta_R} + 
{1\over 2} \left(
   [\epsilon\lambda_L,\theta_L] + [\epsilon\lambda_R,\theta_R]
\right){\partial\over\partial X} 
-  
\nonumber \\   
&\;
- {1\over 6} [\theta_L,[\theta_L,\epsilon\lambda_R]]
{\partial\over\partial\theta_L}
- {1\over 6} [\theta_L,[\theta_R,\epsilon\lambda_L]]
{\partial\over\partial\theta_L} \; +
\nonumber \\   
&\;
+ {1\over 3} [\theta_R,[\theta_L,\epsilon\lambda_L]]
{\partial\over\partial\theta_L}
+ {1\over 3} [\theta_R,[\theta_R,\epsilon\lambda_R]]
{\partial\over\partial\theta_L} \; +
\nonumber \\  
&\; 
- {1\over 6} [\theta_R,[\theta_R,\epsilon\lambda_L]]
{\partial\over\partial\theta_R}
- {1\over 6} [\theta_R,[\theta_L,\epsilon\lambda_R]]
{\partial\over\partial\theta_R} \; +
\label{QBRST} \\   
&\;
+ {1\over 3} [\theta_L,[\theta_R,\epsilon\lambda_R]]
{\partial\over\partial\theta_R}
+ {1\over 3} [\theta_L,[\theta_L,\epsilon\lambda_L]]
{\partial\over\partial\theta_R} \; +
\nonumber \\   
&\;
+ {1\over 24}[\theta_L,[\theta_L,[\theta_R,\epsilon\lambda_L]]] {\partial\over\partial X}
+ {1\over 24}[\theta_L,[\theta_R,[\theta_L,\epsilon\lambda_L]]] {\partial\over\partial X} 
\; +
\nonumber \\   
&\;
+ {1\over 24}[\theta_R,[\theta_L,[\theta_L,\epsilon\lambda_L]]] {\partial\over\partial X}
+ {1\over 24}[\theta_R,[\theta_R,[\theta_R,\epsilon\lambda_L]]] {\partial\over\partial X}
\; +
\nonumber \\   
&\;
+ {1\over 24}[\theta_R,[\theta_R,[\theta_L,\epsilon\lambda_R]]] {\partial\over\partial X}
+ {1\over 24}[\theta_R,[\theta_L,[\theta_R,\epsilon\lambda_R]]] {\partial\over\partial X}
\; +
\nonumber \\   
&\;
+ {1\over 24}[\theta_L,[\theta_R,[\theta_R,\epsilon\lambda_R]]] {\partial\over\partial X}
+ {1\over 24}[\theta_L,[\theta_L,[\theta_L,\epsilon\lambda_R]]] {\partial\over\partial X}\;
+\ldots
\end{align}
\remv{AXIOM}\rem{
delTL := LL - (1/6)*t^2*Nst3(TL,TL,LR) - (1/6)*t^2*Nst3(TL,TR,LL) _
         + (1/3)*t^2*Nst3(TR,TL,LL) + (1/3)*t^2*Nst3(TR,TR,LR)
delTR := LR - (1/6)*t^2*Nst3(TR,TR,LL) - (1/6)*t^2*Nst3(TR,TL,LR) _
         + (1/3)*t^2*Nst3(TL,TR,LR) + (1/3)*t^2*Nst3(TL,TL,LL)
delXX :=  (1/2)*Nst2(LL,TL) + (1/2)*Nst2(LR,TR) _
          + (1/24)*t^2*Nst4(TL,TL,TR,LL) + (1/24)*t^2*Nst4(TL,TR,TL,LL) _
          + (1/24)*t^2*Nst4(TR,TL,TL,LL) + (1/24)*t^2*Nst4(TR,TR,TR,LL) _
          + (1/24)*t^2*Nst4(TR,TR,TL,LR) + (1/24)*t^2*Nst4(TR,TL,TR,LR) _
          + (1/24)*t^2*Nst4(TL,TR,TR,LR) + (1/24)*t^2*Nst4(TL,TL,TL,LR)
compensator := - (1/2)*t^2*Nst2(LR,TL) - (1/2)*t^2*Nst2(LL,TR) 
               + (1/4)*t^4*Nst3(TL,LL,XX) - (1/4)*t^4*Nst3(LL,TL,XX) _
               + (1/4)*t^4*Nst3(TR,LR,XX) - (1/4)*t^4*Nst3(LR,TR,XX) _
               + (-1/24)*t^4*Nst4(TL,TR,TR,LL) + (-1/24)*t^4*Nst4(TR,TL,TR,LL) _
               + (-1/24)*t^4*Nst4(TR,TR,TL,LL) + (-1/24)*t^4*Nst4(TL,TL,TL,LL) _
               + (-1/24)*t^4*Nst4(TR,TL,TL,LR) + (-1/24)*t^4*Nst4(TL,TR,TL,LR) _
               + (-1/24)*t^4*Nst4(TL,TL,TR,LR) + (-1/24)*t^4*Nst4(TR,TR,TR,LR)
}
In this formula, the first line is of the order $1$, and the following 
lines are of the order $R^{-1}$, and the dots stand for the terms of the order
$O(R^{-2})$.
The currents:
\begin{align}
 -J& = dg g^{-1} =  \;  e^{\theta}(de^X e^{-X})e^{-\theta} + de^{\theta} e^{-\theta} =
 \\   
 & =  \; e^{\theta}\left(dX + {1\over 2} [X,dX]\right)e^{-\theta} + 
   d\theta + {1\over 2} [\theta, d\theta] + {1\over 6} [\theta, \theta, d\theta] +
   {1\over 24} [\theta, \theta, \theta, d\theta]
+ \ldots
\nonumber
\end{align}
\begin{align}
-J_{\bar{3}} & = d\theta_L + [\theta_R,dX] + {1\over 6}[\theta,\theta, d\theta]_L + \ldots
\\     
-J_{\bar{2}} & = dX + {1\over 2}[\theta,d\theta]_{\bar{2}} + 
{1\over 2} [\theta, [\theta, dX]]_{\bar{2}} + 
{1\over 24} [\theta,\theta,\theta,d\theta]_{\bar{2}} + \ldots
\\  
-J_{\bar{1}} & = d\theta_R + [\theta_L,dX] + {1\over 6}[\theta,\theta, d\theta]_R + \ldots
\end{align}
The action (\ref{SAdS}) up to the order $R^{-3}$ is:
\begin{equation}\label{ActionToOrderThree}
S = 
 \;\int d^2\tau\; \left[
    R^{-1}\partial_+\theta_R \partial_-\theta_L 
    \;+ \;
    {1\over 2} R^{-2}\partial_+x\partial_-x
\;+\; R^{-3}\left(\;L_3 \;+\; L_4 \;\right)\;+\;\ldots\right]
\end{equation}
where:
\begin{align}
   L_3\;= \;&    - {1\over 2}\left(\;
      [\theta_R,\partial_+\theta_R]\;,\;\partial_-x \;
   \right) 
   - {1\over 2}\left(\;
      \partial_+x\;,\;[\theta_L,\partial_-\theta_L] \;
   \right) 
\end{align}

\begin{align}
   L_4\;=\;&\;
-{1\over 24}([\theta_L,\partial_+\theta_L],[\theta_L,\partial_-\theta_L])
-{1\over 24}([\theta_R,\partial_+\theta_R],[\theta_R,\partial_-\theta_R])\;+
\\    
&\;
-{1\over 12}([\theta_R,\partial_+\theta_R],[\theta_L,\partial_-\theta_L])\;-
\\    
&\;
-{1\over 6}([\theta_R,\partial_+\theta_L],[\theta_R,\partial_-\theta_L])
-{1\over 6}([\theta_L,\partial_+\theta_R],[\theta_L,\partial_-\theta_R])\;-
\\   
&\;
-{1\over 3}([\theta_L,\partial_+\theta_R],[\theta_R,\partial_-\theta_L])
\end{align}
}

\subsubsection{First order formalism}
We get rid of the leading term $R^{-1}\partial_+\theta_R \partial_-\theta_L$ using the first order 
formalism:
\begin{align}
S = \;
 \;\int d^2\tau\;& \left[
    R^{-2}(\tilde{p}_{1+}\partial_-\theta_L) 
    + R^{-2}(\tilde{p}_{3-}\partial_+\theta_R)
    - R^{-3} (\tilde{p}_{1+}\tilde{p}_{3-})    \;+ \;\right.
\nonumber \\   
& +\;\left.
    {1\over 2} R^{-2}\partial_+x\partial_-x
\;+\; R^{-2}\left(\;L_3 \;+\; L_4 \;\right)\;+\;\ldots\right]
\end{align}
where dots stand for the terms of the higher order in $R^{-1}$ (including terms
the order $R^{-3}$, of which the one which depends on $\tilde{p}$, namely  $R^{-3}(\tilde{p}_{1+}\tilde{p}_{3-})$, 
we put explicitly on the first line). Integrating out $\tilde{p}_{\pm}$:  
\begin{equation}
\tilde{p}_{1+} = R\partial_+\theta_R\;,\;\;
\tilde{p}_{3-} = R\partial_-\theta_L
\end{equation}
generates $R^{-1}\partial_+\theta_R \partial_-\theta_L$ and brings us back to (\ref{ActionToOrderThree}).

Importantly, we can remove the leading nonlinear terms $R^{-2}(L_3 + L_4)$ by a
redefinition of $\tilde{p}$. (Otherwize the flat space limit would not have been a 
free theory.) It is done as follows:
\begin{align}
   p_{1+} =\quad & \tilde{p}_{1+} + {1\over 2}[\theta_L,\partial_+x] \;+
\nonumber \\   
+\;& 
{1\over 24}[\theta_L,[\theta_L,\partial_+\theta_L]]
\;+\; {1\over 24}[\theta_L,[\theta_R,\partial_+\theta_R]]\;+
\nonumber \\   
+\;&
{1\over 6}[\theta_R,[\theta_R,\partial_+\theta_L]] 
\;+\; {1\over 6}[\theta_R,[\theta_L,\partial_+\theta_R]]
\\[5pt]
   p_{3-} =\quad & \tilde{p}_{3-} + {1\over 2}[\theta_R,\partial_-x] \;+
\nonumber \\   
+\;& 
{1\over 24}[\theta_R,[\theta_R,\partial_-\theta_R]]
\;+\; {1\over 24}[\theta_R,[\theta_L,\partial_-\theta_L]]\;+
\nonumber \\   
+\;&
{1\over 6}[\theta_L,[\theta_L,\partial_-\theta_R]] 
\;+\; {1\over 6}[\theta_L,[\theta_R,\partial_-\theta_L]]
\end{align}
After these changes of variables, the leading terms in the action are:
\begin{equation}
S = \;
 \;\int d^2\tau\; \left[
    R^{-2}(p_{1+}\partial_-\theta_L) 
    + R^{-2}(p_{3-}\partial_+\theta_R)
    + {1\over 2} R^{-2}\partial_+x\partial_-x
\right]
\end{equation}

\subsection{Relation between $J_{\pm}$ and $d_{\pm}$}\label{sec:RelationJAndD}
We observe that in the flat space limit $J_{3-}$ and $J_{1+}$ go like $R^{-3/2}$. 
We identify:
\begin{align}
   J_{1+} =&\; -d_+ + O(R^{-5/2})
\\     
   J_{3-} =&\; -d_- + O(R^{-5/2})
\end{align}
In terms of $x$ and $\theta$, at the order $R^{-3/2}$:
\begin{align}
J_{1+} =\;& 
-\partial_+\theta_R - [\theta_L,\partial_+ x]\; -
\nonumber \\   
&
-{1\over 6}[\theta_L,[\theta_L,\partial_+\theta_L]]
-{1\over 6}[\theta_R,[\theta_R,\partial_+\theta_L]]\; =
\nonumber \\   
=\;&
-p_{1+} - {1\over 2}[\theta_L,\partial_+ x]\; -
{1\over 8}[\theta_L,[\theta_L,\partial_+\theta_L]]
\end{align}

\subsection{Global symmetry currents}\label{sec:GlobalCurrents}
The matter contribution into the global symmetry currents:
\begin{align}
-j_+  = &\; g^{-1}\left( 
      J_{\bar{3}+} + 2 J_{\bar{2}+} + 3 J_{\bar{1}+} 
   \right) g
\\   
j_-  = &\; g^{-1}\left( 
      3J_{\bar{3}-} + 2 J_{\bar{2}-} +  J_{\bar{1}-} 
   \right) g
\end{align}
For example consider the global symmetry currents $j_{3+}$ and $j_{3-}$. 

\vspace{10pt}
\noindent Up to $O(R^{-7/2})$ and up to terms which do not contain $\partial_+\theta_R$:
\begin{align}
   j_{3+} = &\; \partial_+\Psi_3 + 4[\partial_+\theta_R,x] + 2[\theta_L,[\theta_L,\partial_+\theta_R]]
+ {2\over 3} [\theta_R,[\theta_R,\partial_+\theta_R]] + \ldots\; 
\nonumber \\   
\mbox{\tt \small where }
\Psi_3 = &\;
   \theta_L + [x,\theta_R] \; - {1\over 3} [\theta_L,[\theta_L,\theta_R]]
\end{align}
Up to $O(R^{-5/2})$:
\begin{align}
j_{3-} = &\; \partial_-\Psi_3 
- 4\partial_-\theta_L - {2\over 3}[\theta_L,[\theta_L,\partial_-\theta_R]] + \ldots\; = 
\nonumber \\   
= &\; \partial_-\Psi_3
- 4d_{3-} + 4[\theta_R,\partial_-x] + {2\over 3}[\theta_R,[\theta_R,\partial_-\theta_R]]\;+\ldots
\end{align}
\vspace{10pt}
\noindent Similarly:
\begin{align}
j_{1+} = &\; \partial_+\Psi_1
+ 4d_{1+} - 4[\theta_L,\partial_+x] - {2\over 3}[\theta_L,[\theta_L,\partial_+\theta_L]]\;+\ldots
\\     
j_{1-} = &\; \partial_- \Psi_1
- 4[\partial_-\theta_L,x] - 2[\theta_R,[\theta_R,\partial_-\theta_L]]
- {2\over 3} [\theta_L,[\theta_L,\partial_-\theta_L]]\; + \ldots
\end{align}
where $\Psi_1$ is given by (\ref{QandPsi}).
The density of a local conserved charge is defined up to a total derivative.

Therefore, let us redefine $j_{\pm} \to S_{\pm}$, by removing total derivatives: 
\begin{align}
j_{3\pm} = &\; \partial_{\pm}\Psi_3 + S_{3\pm}
\nonumber \\  
j_{1\pm} = &\; \partial_{\pm}\Psi_1 + S_{1\pm}
\end{align}
In the flat space expansion:
\begin{equation}
S_{1+}\simeq R^{-3/2},\;S_{1-} \simeq R^{-5/2},\; S_{3-}\simeq R^{-3/2},\;
S_{3+}\simeq R^{-5/2}
\end{equation}
We should identify $S_{1+}$ and $S_{3-}$ with the supersymmetry currents of the flat 
space superstring. Explicitly we have:
\begin{align}
   S_{1+} =\;& 4\left(
      p_{1+} - {1\over 2}[\theta_L,\partial_+x] 
      - {1\over 24}[\theta_L,[\theta_L,\partial_+\theta_L]]
   \right)\; = 
\nonumber \\
=\;& 4\left(
   d_{1+} - [\theta_L,\partial_+x] 
   - {1\over 6}[\theta_L,[\theta_L,\partial_+\theta_L]]
\right)
\label{S1Plus}
\\   
  - S_{3-} =\;& 4\left(
      p_{3-} - {1\over 2}[\theta_R,\partial_-x] 
      - {1\over 24}[\theta_R,[\theta_R,\partial_-\theta_R]]
   \right)\; =
\nonumber \\   
=\;& 4\left(
   d_{3-} - [\theta_R,\partial_-x] 
   - {1\over 6}[\theta_R,[\theta_R,\partial_-\theta_R]]
\right)
\label{S3Minus}
\end{align}

\paragraph     {$U_{\rm AdS}$ in terms of the global currents:}
Now we can write Eq. (\ref{UAdSviaD}) precisely, including the terms with $\theta$:
\begin{equation}\label{CompleteURR5Form}
U_{\rm AdS} = \mbox{Str}(S_{1+} S_{3-})
\end{equation}

\subsection{Unphysical vertex of the order $R^{-3}$}
Let us consider the following example of the unphysical vertex:
\begin{equation}\label{UZero}
U_{\bar{0}} = [j_{\bar{1}+},j_{\bar{3}-}] + [j_{\bar{3}+},j_{\bar{1}-}] =
[S_{\bar{1}+},S_{\bar{3}-}] \simeq R^{-3}
\end{equation}
In this case the flat space limit of the unphysical vertex appears to be 
perfectly physical, and in fact corresponds to turning on the constant RR 
3-form field strength. Indeed, there is a term of the type $d_+d_-$ 
plus terms containing $\theta$'s:
\begin{equation}
U_{\bar{0}} = [d_+,d_-] + \ldots
\end{equation}
A careful analysis of the index structure shows that this actually
corresponds to the constant RR 3-form field strength.

The flat space limit of the vertex operator for the beta-deformation is
generally speaking of the order $\varepsilon R^{-3}$. It typically starts with $xdx\wedge dx$,
plus terms of the type $d_+d_-$ (which are also of the order $R^{-3}$, since $d_{\pm}$ 
are of the order $R^{-3/2}$). Plus terms with $\theta$. The leading bosonic 
term $xdx\wedge dx$ describes a NSNS $B_{NSNS}$-field. At the order $\varepsilon R^{-3}$ we can 
only see the constant NSNS field strength $H_{NSNS}$. The terms with $d_+d_-$ 
describe the constant RR field strength $H_{RR}$. We conclude that we see some 
constant $H_{NSNS}$ and some constant $H_{RR}$. This is nice.

But let us expand it at a different point in AdS, the point at which the
field strengths are zero. Then the leading terms in the vertex will be of 
the order $R^{-4}$.

\subsection{Unphysical vertex of the order $R^{-4}$}
\subsubsection{Definition of the vertex and how the descent procedure does not work}
Consider another example of the unphysical vertex: 
\begin{align}
U_{\bar{2}} =\;& \;{1\over 2}\;\mbox{Str}\left(
   \;[B_2,j_{1}]\wedge j_{1}\; + \;[B_2,j_{3}]\wedge j_{3}\;
\right)\;=
\label{UTwo}\\    
=\;& \mbox{Str}\left(\;B_{\bar{2}}\;[j_{\rm odd}\;,\;j_{\rm odd}]\;\right)
\end{align}
The flat space limit of an unintegrated unphysical vertex was derived in \cite{Bedoya:2010qz}:
\begin{equation}
V_{\bar{2},\rm \; flat} = 
[[\theta_R,[\theta_R,\epsilon\lambda_R]],[\theta_R,[\theta_R,\epsilon\lambda_R]] + 
[[\theta_L,[\theta_L,\epsilon\lambda_L]],[\theta_L,[\theta_L,\epsilon\lambda_L]] 
\end{equation}
What happens if we apply to it the flat space descent procedure? Observe:
\begin{equation}
\partial_- [\theta_R,[\theta_R,\epsilon\lambda_R]] = Q (3S_{3-})
\end{equation}
\ifodd\amshow
%{\tt \href{notes_extra.pdf#href:DescentOfTLTLLL}{Calculation} }
\fi
Notice that in flat space the supersymmetry current $S_{3-}$ is holomorphic.
Therefore the second step of the descent procedure is zero:
\begin{equation}
\partial_+[\;[\theta_R,[\theta_R,\epsilon\lambda_R]] \;,\; S_{3-}\;] = 0
\end{equation}
This means that the corresponding integrated vertex, defined by the descent 
procedure, is zero. 
(If it were not zero, it would have been of the order $R^{-3}$.)
\paragraph     {Conclusion:}The leading flat space limit of (\ref{UTwo}) is {\em not} 
related to $V_{\bar{2},\; \rm flat}$ by a descent procedure.  

\subsubsection{Explicit formula for the vertex in flat space}
We observe:
\begin{align}
   \epsilon Q\int U_{\bar{2}} = - \int \mbox{Str}\left(
      \;
      \left[\;B_2\;,\; g^{-1}(\epsilon\lambda_L - \epsilon\lambda_R)g\;\right]
      \;(dj_1 + dj_3)
      \;
   \right)
\end{align}
The variation is proportional to the equation of motion $dj_1 = 0$, $dj_3 = 0$. To 
compensate this variation we need the field redefinition:
\begin{equation}
\epsilon Q_1 = 4\left[
   \;B_2\;,\;g^{-1}(\epsilon\lambda_L - \epsilon\lambda_R)g\;
\right]_3^{\alpha}t_{\alpha}^3
\;+\;
4\left[
   \;B_2\;,\;g^{-1}(\epsilon\lambda_L - \epsilon\lambda_R)g\;
\right]_1^{\hat{\alpha}}t_{\hat{\alpha}}^1
\end{equation}
Then the deformed action:
\begin{equation}
S_{\rm AdS} + \int \mbox{Str}\left(\;B_2\;j_{\rm odd}\wedge j_{\rm odd}\;\right)
\end{equation}
is invariant under the deformed BRST transformation $\epsilon (Q + Q_1)$. 

To get the expression starting with $R^{-4}$, we do the field redefinition with 
the vector field $K$ given by (\ref{DefKLeft}) plus (\ref{DefKRight}). Then the deformed action
\begin{equation}
S_{\rm AdS} + KS_{\rm AdS} + 
\int \mbox{Str}\left(\;B_2\;j_{\rm odd}\wedge j_{\rm odd}\;\right)
\end{equation}
is invariant under the deformed BRST transformation:
\begin{align}
& \epsilon Q + \epsilon Q'_1
\nonumber \\    
\mbox{\tt \small where }&
\epsilon Q'_1 = \epsilon Q_1 + [K,\epsilon Q] = 
\nonumber \\   
& \qquad =
-{16\over 3} [B_2,\;[\theta_L,[\theta_L,\epsilon\lambda_L]]]^{\alpha} t^3_{\alpha}
+{16\over 3} [B_2,\;[\theta_R,[\theta_R,\epsilon\lambda_R]]]^{\hat{\alpha}} 
t^1_{\hat{\alpha}}
\end{align}
Using (\ref{KSAdS}) we get:
\begin{align}
& S_{\rm AdS} + KS_{\rm AdS} 
+ \int \mbox{Str}\left(\;B_2\;j_{1}\wedge j_{1}\;\right)
+ \int \mbox{Str}\left(\;B_2\;j_{3}\wedge j_{3}\;\right)
= 
\nonumber \\   
=\;& S_{\rm AdS} +\left(
-  \int d^2\tau\;\mbox{Str}
\left(
   \partial_+[B_2,\Psi_1]\; j_{1-} - \partial_-[B_2,\Psi_1]\; j_{1+}
\right)\; \right.+
\nonumber \\  
&
\phantom{S_{\rm AdS}\qquad }\left.
 + \int d^2\tau\;\mbox{Str}([B_2,j_{1+}]\;j_{1-})\;+\; (1\rightarrow 3)
\right)\; =
\nonumber \\     
=\;& S_{\rm AdS} 
+\left( \int d^2\tau\;\mbox{Str}
\left(
   \;\left[B_2\;,\;(j_{1+} - \partial_+\Psi_1)\right]
   \;(j_{1-} - \partial_-\Psi_1)\;
\right) + (1\rightarrow 3)\right)\; =
\nonumber \\    
=\;& S_{\rm AdS} 
+ \int d^2\tau\;\mbox{Str}
\left(
   B_2\left[S_{1+},
      S_{1-}\right] +
   B_2\left[S_{3+},
      S_{3-}
   \right]
\right) 
\end{align}
Now formulas of Section \ref{sec:GlobalCurrents} imply that the flat space limit is 
of the order $R^{-4}$:
\begin{align}
 U_{\bar{2},\rm \; flat} \;=\;\mbox{Str}\left(\phantom{+} B_2
\vphantom{\int}\right. & 
    \left[ 4 [d_{1+},x] + 2[\theta_L,[\theta_L,d_{1+}]]
       + {2\over 3} [\theta_R,[\theta_R,d_{1+}]]
    \right.\;,\;
\nonumber \\   
&\left.
   - 4d_{3-} + 4[\theta_R,\partial_-x] 
   + {2\over 3}[\theta_R,[\theta_R,\partial_-\theta_R]]\;
\right]+
\nonumber \\   
+\;B_2&\left[ 4d_{1+} - 4[\theta_L,\partial_+x] 
   - {2\over 3}[\theta_L,[\theta_L,\partial_+\theta_L]]
   \right.\;,\;
\\    
&\left. - 4 [d_{3-},x]  - 2 [\theta_R,[\theta_R,d_{3-}]]
- {2\over 3}[\theta_L,[\theta_L,d_{3-}]]
\right]
\left.\vphantom{\int}\right)
\end{align}
where $\ldots$ stand for the terms of the same order $R^{-4}$ containing higher number
of thetas. Also the ghosts contribute:
\begin{align}
U_{\bar{2},\;\rm flat, \; gh} = 4\left[
   [\theta_L,\{w_{1+},\lambda_L\}] \;,\; S_{3-}
\right]
\end{align}
but their contribution will not be very important here. 

We observe that there is the term $xd_+d_-$, more precisely:
\begin{equation}\label{Bxdd}
   16 \; \mbox{Str}\left(
      [B_2,x][d_{1+},d_{3-}]
   \right)
\end{equation}
which usually corresponds to the Ramond-Ramond field. Since it is odd under
the worldsheet parity ({\it i.e.} under the exchange $d_+\leftrightarrow d_-$) we should
have concluded that it corresponds to the Ramond-Ramond 3-form field strength
$H$. But we also find that $dH\neq 0$. In the usual notations (\ref{Bxdd}) would 
correspond to $H = \iota_{B_2\wedge x}F$, where $F$ is the leading flat space limit of the 
RR field of $AdS_5\times S^5$. This is not a closed form. Naively this is in 
contradiction with \cite{Berkovits:2001ue}, as $dH=0$ is one of the SUGRA equations of motion. 
The resolution is, as explained in Section \ref{sec:WildDeformations}, that $U_{\bar{2},\;\rm flat}$ is actually not 
annihilated by $Q_{\rm flat}$. 

\subsection{Demonstration of the LHS of (\ref{DefUB}) being nonzero.}
Let us calculate the variation of the AdS action along the vector field (\ref{DeformQ}).
We get the following expression of the order $R^{-4}$:
\begin{align}
   &
   \left(
      [B_2,[\theta_L,[\theta_L,\epsilon\lambda_L]]]^{\hat{\alpha}}\; 
      t_{\hat{\alpha}}^1
   \right)
   \;\;S_{\rm AdS} \;=
   \\    
   =\;& \int d^2\tau \;
   \mbox{Str}\left(
      \partial_-[\theta_L,[\theta_L,\epsilon\lambda_L]]\;S_{1+} -
      \partial_+[\theta_L,[\theta_L,\epsilon\lambda_L]]\;S_{1-}
   \right)
\end{align}
The term with $\partial_-[\theta_L,[\theta_L,\epsilon\lambda_L]]\;S_{1+}$ generates:
\begin{align}
   \int\;d^2\tau\; \mbox{Str}\left(
      [d_{3-},[\theta_L,\epsilon\lambda_L]]\;d_{1+}
      + [\theta_L,[d_{3-},\epsilon\lambda_L]]\;d_{1+}
   \right)
\end{align}
which does not have anything to cancel with. This demonstrates that the LHS 
of (\ref{DefUB}) is nonzero. 

\subsection{Parity even physical vertex}
It is also interesting to consider the following {\em physical} vertex:
\begin{equation}\label{UTwoPhys}
U_{\bar{2},\;\rm phys}= {1\over 2}\;\mbox{Str}\left(
   \;[B_2,j_{1}]\wedge j_{1}\; - \;[B_2,j_{3}]\wedge j_{3}\;
\right)
\end{equation}
It differs from (\ref{UTwo}) by the relative sign of the two terms.
Unlike (\ref{UTwo}), this vertex does satisfy the physical condition (\ref{PhysicalCondition}), and
does correspond to a meaningful excitation of $AdS_5\times S^5$. Notice that $U_{\bar{2},\;\rm phys}$
is parity-even, therefore it should correspond to either a metric, or a 
dilaton, or a RR 1-form, or a RR 5-form. 

As becomes clear from Section \ref{sec:NormalForm}, the flat space limit of the parity 
even vertex is the linear dilaton background. (Whereas the parity odd 
vertex is unphysical and does not correspond to anything.)

\section{Bringing the action to the normal form of \cite{Berkovits:2001ue}}\label{sec:NormalForm}
This section was added in the revised version of the paper. 

Generally speaking, given a sigma-model, we can always rewrite it in many 
different forms using field redefinitions, introducing Lagrange multipliers,
alternative gauge fixings, {\it etc.}. 
In order to make contact with the spacetime description in terms of Type IIB
SUGRA fields, the authors of \cite{Berkovits:2001ue} used a special ``normal form'' of the 
sigma-model action. The definition of this normal form depends on how the
BRST symmetry acts. Although in our case the action of the sigma-model does
not change, but the BRST operator does get deformed. Therefore, the 
{\em normal form} of the action does get deformed. We will now study the 
deformation of the normal form of the action. We will show that it leads
to the nontrivial spin connection. It turns out that the vector components of 
the left and right spin connections do not coincide (contrary to what was 
conjectured in \cite{Berkovits:2001ue}); this is why the deformation is nonphysical.

We will use the notations of \cite{Berkovits:2001ue}; we also recommend \cite{Guttenberg:2008ic} for the detailed 
explanations of the formalism. We will continue using the flat space 
notations (with $\Gamma$-matrices) and the AdS notations (commutators and $\mbox{Str}$) 
intermittently, as explained in Section \ref{sec:AdSNotations}. 

\subsection{Action in terms of $d_{\pm}$}
As we explained, the action is undeformed:
\begin{align}
S_{\rm flat} = \;
\int d\tau^+ d\tau^- & \;\left[
   {1\over 2}\partial_+x^m\partial_-x^m + p_{\alpha +}\partial_-\theta^{\alpha}_L 
   + p_{\hat{\alpha}-}\partial_+\theta^{\hat{\alpha}}_R \;+ 
\right. 
\label{SFlatTraditional} \\   
&\;\;\; 
\left. 
   +\; w_{\alpha +}\partial_-\lambda^{\alpha}_L 
   + w_{\hat{\alpha}-}\partial_+\lambda^{\hat{\alpha}}_R
\right]\; =
\\    
 = \int d\tau^+ d\tau^- & \;\mbox{Str }\left[
   {1\over 2}\partial_+x_2\partial_-x_2 + p_{1+}\partial_-\theta_L +
   p_{3-}\partial_+\theta_R \;+ 
\right.
\nonumber \\   
&\;\;\; 
\left. \qquad
   +\; w_{1+}\partial_-\lambda_L + w_{3-}\partial_+\lambda_R
\right]
\label{SFlatStr}
\end{align}
(Eq. (\ref{SFlatTraditional}) uses traditional notations, while Eq. (\ref{SFlatStr}) uses AdS notations.)
The deformation only touches the BRST operator. In order to bring the action
to the form of \cite{Berkovits:2001ue}, we need to trade $p_{\pm}$ for $d_{\pm}$, where $d_{\pm}$ is defined as the 
density of the BRST charge: 
\begin{equation}\label{NormalBRST}
Q_{L|R} = \oint \lambda_{L|R}d_{\pm}
\end{equation}
In the undeformed theory, the relation between $d_{\pm}$ and $p_{\pm}$ is given by 
Eqs. (\ref{S1Plus}), (\ref{S3Minus}):
\begin{align}
   p_{1+} = \;& d_{1+} - {1\over 2} [\theta_L,\;\partial_+x]
   - {1\over 8} [\theta_L,\;[\theta_L,\;\partial_+\theta_L]]
   \\       
   p_{3-} = \;& d_{3-} - {1\over 2} [\theta_R,\;\partial_-x]
   - {1\over 8} [\theta_R,\;[\theta_R,\;\partial_-\theta_R]]
\end{align}
After the deformation, this relation is modified. Let us consider the case
when $B_R=0$ (only the left deformation):
\begin{align}
   p_{1+} = \;& d_{1+} - {1\over 2} [\theta_L,\;\partial_+x]
   - {1\over 8} [\theta_L,\;[\theta_L,\;\partial_+\theta_L]] \;+
   \nonumber \\
   & + [\theta_L,[\theta_L,\;[B_2,\;S_{1+}]]]
   \label{P1PlusDeformed}\\       
   p_{3-} = \;& d_{3-} - {1\over 2} [\theta_R,\;\partial_-x]
   - {1\over 8} [\theta_R,\;[\theta_R,\;\partial_-\theta_R]]
\end{align}
Let us substitute $S_{1+}$ from (\ref{S1Plus}) into (\ref{P1PlusDeformed}): 
\begin{align}
   p_{1+} = \;& d_{1+} - {1\over 2} [\theta_L,\;\partial_+x]
   - {1\over 8} [\theta_L,\;[\theta_L,\;\partial_+\theta_L]] \;+
   \nonumber \\
   & + 4\;\left[\theta_L,\;\left[\theta_L,\;\left[
            B_2\;,\;\left(
               d_{1+} - [\theta_L,\;\partial_+x] - 
               {1\over 6}[\theta_L,\;[\theta_L,\;\partial_+\theta_L]]
               \right)
      \right]\right]\right]
   \label{DeformedP1Expanded}
\end{align}
Therefore, we get the following formula for the action, which at this point is
almost in the normal form of \cite{Berkovits:2001ue}:
\begin{align}
S = \;& \int d\tau^+ d\tau^- \;\mbox{Str}\;\left({1\over 2} 
\partial_+ x_2 \partial_- x_2 + d_{1+}\partial_-\theta_L + 
d_{3-}\partial_+\theta_R \;-\right.
\nonumber \\  
\;& \;\;
- {1\over 2} [\theta_L,\;\partial_+x]\partial_-\theta_L
- {1\over 8} [\theta_L,\;[\theta_L,\;\partial_+\theta_L]]\partial_-\theta_L \;-
\nonumber \\  
\;& \;\;
- {1\over 2} [\theta_R,\;\partial_-x]\partial_+\theta_R
- {1\over 8} [\theta_R,\;[\theta_R,\;\partial_-\theta_R]]\partial_+\theta_R\;+
\label{ActionNormalFormNoOmega} \\  
\;& \;\; +
w_{1+}\partial_-\lambda_3 + w_{3-}\partial_+\lambda_R \;+
\nonumber \\  
\;& \left.\;\; +\; 4\;\left[
            B_2\;,\;\left(
               d_{1+} - [\theta_L,\;\partial_+x] - 
               {1\over 6}[\theta_L,\;[\theta_L,\;\partial_+\theta_L]]
               \right)
      \right]\;
   \left[\theta_L,\;\left[\theta_L,\;\partial_-\theta_L\right]\right]
\right)
\nonumber
\end{align}

\subsection{$B$-field.} 
In particular this allows us to read the $B$-field part:
\begin{align}
B_{MN}\; dZ^M \wedge dZ^N = \mbox{Str}\Big(
- {1\over 2} [\theta_L,\;d x_2] d\theta_L
- {1\over 8} [\theta_L,\;[\theta_L,\; d\theta_L]]\;d\theta_L \;-
\nonumber \\  
 - {1\over 2} [\theta_R,\;d x_2] d\theta_R
- {1\over 8} [\theta_R,\;[\theta_R,\; d\theta_R]]\;d\theta_R \;-
\nonumber \\  
-\; 4\;\left[
            B_2\;,\;\left(
               [\theta_L,\;dx_2] +
               {1\over 6}[\theta_L,\;[\theta_L,\;d\theta_L]]
               \right)
      \right]\;
\left[\theta_L\;,\;\left[\theta_L,\;d\theta_L\right]\right]
\Big)
\end{align}
The 3-form field strength $H = dB$ is:
\begin{align}
H= \;& \mbox{Str}\Big( 
-{1\over 2} [d\theta_L\;,\;dx_2]d\theta_L 
+ {1\over 4} [d\theta_L\;,\;d\theta_L]\;[\theta_L\;,\;d\theta_L] \;-
\nonumber \\  
\;& \qquad -{1\over 2} [d\theta_R\;,\;dx_2]d\theta_R 
+ {1\over 4} [d\theta_R\;,\;d\theta_R]\;[\theta_R\;,\;d\theta_R] \;-
\nonumber \\  
& -\; 4\;\left[
            B_2\;,\;\left(
               [d\theta_L,\;dx_2] +
               {1\over 4}[\theta_L,\;[d\theta_L,\;d\theta_L]]
               \right)
      \right]\;
\left[\theta_L\;,\;\left[\theta_L,\;d\theta_L\right]\right]
\Big)
\nonumber \\  
& +\; 6\;\left[
            B_2\;,\;\left(
               [\theta_L,\;dx_2] +
               {1\over 6}[\theta_L,\;[\theta_L,\;d\theta_L]]
               \right)
      \right]\;
\left[\theta_L\;,\;\left[d\theta_L,\;d\theta_L\right]\right]
\Big)
\label{HField}
\end{align}
For example, let us demonstrate that: 
\begin{equation}
H_{\alpha\beta m}\lambda^{\alpha}\lambda^{\beta} = 0
\label{HLambdaLambda}
\end{equation}
in accordance with \cite{Berkovits:2001ue}. The last row in (\ref{HField}) does not contribute, because 
$\{\lambda_L\;,\;\lambda_L\} = 0$. In the previous rows, the terms containing $dx d\theta_L d\theta_L$ 
combine into:
\begin{align}
{1\over 2}   \mbox{Str}\Big( 
    dx_2\; \Big[\; 
         d\theta_L - 4 [B_2,[\theta_L,[\theta_L,d\theta_L]]]
      \;\;,\;\;
      d\theta_L - 4 [B_2,[\theta_L,[\theta_L,d\theta_L]]]
\;\Big]\Big)
\end{align}
Notice that $\epsilon Q\theta_L = \epsilon \lambda_L + 4\;[B_2,[\theta_L,[\theta_L,\epsilon\lambda_L]]]$ and (\ref{HLambdaLambda}) follows. 

\subsection{Torsion.}
The action (\ref{ActionNormalFormNoOmega}) is almost in the normal form, but not completely. To complete
the procedure described in \cite{Berkovits:2001ue} we have to eliminate some components of the
torsion, namely $T_{\alpha\beta}{}^{\gamma}$. Let us therefore study the torsion.

The 16-beins $E^{\alpha}$ and $E^{\hat{\alpha}}$ are defined as the coefficients of $d_{\pm}$ in the
worldsheet action (\ref{ActionNormalFormNoOmega}):
\begin{align}
 E^{\alpha} =  E^{\alpha}_M dZ^M = \;& d\theta_L^{\alpha} -
   4\;[B_2\;,\;[\theta_L\;,\;[\theta_L\;,\;d\theta_L]]]^{\alpha}
   \\   
 E^{\hat{\alpha}} =  E^{\hat{\alpha}}_M dZ^M = \;& d\theta_R^{\hat{\alpha}}
\end{align}
Notice that the pure spinor terms in the action (\ref{ActionNormalFormNoOmega}) are the same as in flat 
space, therefore $\Omega_M{}^{\alpha}_{\beta} = \widehat{\Omega}_M{}^{\hat{\alpha}}_{\hat{\beta}} = 0$. Therefore the torsion is defined as in 
flat space: $T^{\alpha} = T^{\alpha}_{MN} dZ^M dZ^N = dE^{\alpha}$, $T^{\hat{\alpha}} = T^{\hat{\alpha}}_{MN} dZ^M dZ^N = dE^{\hat{\alpha}}$.
In particular:
\begin{align}
   T^{\alpha} =\;& 
   -6\;[B_2\;,\;[\theta_L\;,\;[d\theta_L\;,\;d\theta_L]]]
   \\  
\mbox{ \tt\small in other words }   
T_{\alpha\beta}^{\gamma} =\;& -6\;\Gamma_{\alpha\beta}^n
   (\overline{B}^m_2\Gamma_m \Gamma_n\theta_L)^{\gamma}
\end{align}
Here the notation $\overline{B}^m$ stands for: $B^m$ for $m\in \{0,1,\ldots 4\}$ and $-B^m$ 
for $m\in \{5,\ldots,9\}$. The difference between $B$ and $\overline{B}$ does not play any role
in our discussion here; it is an artifact of notations in Section \ref{sec:AdSNotations}. 

\paragraph     {Removing $T^{\gamma}_{\alpha\beta}$.}
As instructed in \cite{Berkovits:2001ue}, we have to remove $T^{\gamma}_{\alpha\beta}$ by a special field redefinition
which at the same time modifies the spin connection $\Omega_{\alpha}^{[mn]}$ and $\Omega_{\alpha}^{(s)}$.
This is done in the following way. Notice that the following {\em field  
redefinition} $d\to \tilde{d}$, parametrized by $h^{a\alpha}(Z)$:
\begin{align}
   d_{\alpha +} =\;& \tilde{d}_{\alpha +} + 
   h^{b\beta} \Gamma_{\alpha\beta}^k
   (w_{+}\Gamma_b\Gamma_k\lambda)
\end{align}
does not change the expression (\ref{NormalBRST}) for the BRST current, and therefore is a
residual field redefinition preserving the normal form of \cite{Berkovits:2001ue} of the 
worldsheet action/BRST structure. This field redefinition changes the string
worldsheet action by adding to it the term:
\begin{equation}
\partial_- Z^M E_M^{\alpha} \Gamma^k_{\alpha\beta} h^{b\beta} 
(w_{+}\Gamma_b\Gamma_k\lambda)
\end{equation}
which encodes the modification of the left connection $\Omega_{\alpha}$: 
\begin{equation}\label{OmegaFromH}
\Omega_{\alpha}^{(s)}  = \Gamma^k_{\alpha\beta} h^{k\beta} \;\;,\quad
\Omega_{\alpha}^{[mn]} = \Gamma^{[m}_{\alpha\beta} h^{n]\beta}
\end{equation}
This changes the $T_{MN}^{\alpha}$:
\begin{align}
   T_{MN}^{\alpha} \to T_{MN}^{\alpha} + 2E_{(M}^{\beta} \Omega_{N)}{}^{\alpha}_{\beta}
= T_{MN}^{\alpha} + E^{\alpha'}_M E^{\beta'}_N \Gamma_{\alpha'\beta'}^b
\Gamma^b_{\gamma'\delta'} \Gamma_c^{\alpha\gamma'} h^{c\delta'}
\end{align}
Taking $h^{a\alpha}$ as follows:
\begin{equation}
   h^{a\alpha} = 6 \overline{B}_2^a\theta_L^{\alpha}
\end{equation}
we get rid of $T_{\alpha\beta}^{\gamma}$ ({\it i.e.} the $T^{\gamma}_{\alpha\beta}$ calculated with this new $\Omega$ is zero) 
at the price of generating $\Omega^{(s)}_{\alpha}$ and $\Omega^{[mn]}_{\alpha}$ given by (\ref{OmegaFromH}). 
\ifodd\amshow
\\ \includegraphics{snapshots/removing-torsion.png} \\
\fi
Notice that $\Omega_{\hat{\alpha}}^{(s)} = 0$, as it should be. Also notice that the right connection 
remains zero, both $\hat{\Omega}_{\hat{\alpha}}^{(s)}$ and  $\hat{\Omega}_{\alpha}^{(s)}$. According to \cite{Berkovits:2001ue} we should then be able to
solve the equations $(D_{\alpha} +\Omega_{\alpha}^{(s)})\Phi = 0$ and $(D_{\hat{\alpha}} +\hat{\Omega}_{\hat{\alpha}}^{(s)})\Phi = 0$ which imply:
\begin{align}
\left(
   {\partial\over\partial\theta_L^{\alpha}} + 
   \Gamma^m_{\alpha\beta}\theta_L^{\beta} {\partial\over\partial x^{m}} +
   6 \overline{B}_2^m\Gamma^m_{\alpha\beta}\theta_L^{\beta} 
\right) \Phi =\;& 0
\label{LeftDerivativeOfPhi}
\\    
\left(
   {\partial\over\partial\theta_R^{\hat{\alpha}}} + 
   \Gamma^m_{\hat{\alpha}\hat{\beta}}\theta_R^{\hat{\beta}} 
   {\partial\over\partial x^m} 
\right) \Phi =\;& 0
\label{RightDerivativeOfPhi}
\end{align}
The first of these equations can be solved by the {\em linear dilaton}\footnote{It is not surprizing that the linear dilaton is involved. In the case of 
bosonic string, also the linear dilaton background does not deform the 
worldsheet action
on a flat worldsheet, but does deform the BRST trasnformation. We would like
to thank Nathan Berkovits for suggesting to look at it from this angle.}:
\begin{equation}
   \Phi = -6 \overline{B}_2^m x^m + \mbox{const}
\end{equation}
but this {\bf does not satisfy the second equation} (\ref{RightDerivativeOfPhi}). In fact, (\ref{RightDerivativeOfPhi}) 
immediately implies that $\Phi = \mbox{const}$. This result can be also formulated in 
the following way:
\begin{itemize}
\item it is not true in this case that $\Omega_m^{(s)} = \widehat{\Omega}_m^{(s)}$
\end{itemize}
Notice that the equality of the vector component of the left and right spin
connections was only conjectured (but not proven) in \cite{Berkovits:2001ue}; our construction 
provides a counter-example to this conjecture. 

We feel that this problem only arizes for the states of low momentum, although
it is not very clear what ``low momentum'' would mean in a generic background.
Perhaps the non-physical vertex only exists in AdS and flat space, and the
corresponding deformation is obstructed at the higher orders of the 
deformation parameter. In any case, as was demonstrated in \cite{Bedoya:2010qz}, the 
non-physical vertices go away if, in addition to the BRST invariance, we also 
impose the 1-loop conformal invariance. This suggests that a modification of 
the BRST complex, taking into account the additional structure provided by 
the $b$-ghost \cite{Nelson:1988ic,Distler:1990ea,Witten:1992yj}, would take care of the problem.

\appendix

\section{Vector field $Y_L$}\label{sec:AppendixConfDimOne}
\subsection{Ansatz for $y_+$}
It is usually assumed that the pure spinor BRST cohomology at the positive 
conformal dimension is trivial. We do not have a general proof of this fact.
Let us consider a particular example which we needed in Section \ref{sec:SpecialCaseB}:
\begin{align}
   Q_{\rm flat} M_+ =\;& 0
\nonumber\\    
\mbox{\tt\small where }M_+ =\;& 
   (\theta\Gamma^m\lambda)(\theta\Gamma_m)_{\alpha}B_L^{\alpha\beta} S_{\beta+} 
- \partial_+(\epsilon' W_L)
\end{align}
We want to prove that exists such $y_+$ that $M_+ = Q_{\rm flat}y_+$. 
{\bf We do not have the complete proof}, but only a schematic expression:
\begin{align}\label{AnsatzForYPlus}
   y_+ = [\theta_L\theta_L N_+] + [\theta_L\theta_L\theta_Ld_+] + 
   [\theta_L^5\partial_+\theta_L] + [\theta_L^4\partial_+x]
\end{align}
where $N_{[mn]+} = (\lambda_L\Gamma_{mn}w_+)$ is the contributions  of the pure spinors to the 
Lorentz current. The term with $[\theta_L\theta_LN_+]$ is necessary because $S_{\beta+}$ contains 
$d_{\beta +}$, and its coefficient in $M_+$ (which is $(\theta_L\Gamma^m\lambda_L)(\theta_L\Gamma_m)_{\alpha}B_L^{\alpha\beta}$) is not 
$Q_{\rm flat}$-exact. Such term can only come from the BRST variation of something
of the type $[\theta_L\theta_LN_+]$. In the next Section we will discuss the structure of 
this term.

\subsection{The term $\theta\theta N_+$}
In order to obtain the term $(\theta_L\Gamma^m\lambda_L)(\theta_L\Gamma_m)_{\alpha}B_L^{\alpha\beta}d_+$, we need the first
term $[\theta_L\theta_LN_+]$ in (\ref{AnsatzForYPlus}) of the form:
\begin{equation}
   [\theta_L\theta_LN_+] \simeq  
   B_{lmnpq} (\theta_L\Gamma^{lmn}\theta_L)(\lambda_L\Gamma^{pq}w_+)
\end{equation}
where $B_{lmnpq}$ is a self-dual antisymmetric tensor defined so that:
\begin{equation}
   B_{lmnpq}\Gamma^{\alpha\beta}_{lmnpq} = B^{\alpha\beta}
\end{equation}
We observe that $Q_{\rm flat}$ of so defined $[\theta_L\theta_LN_+]$ does not contain $w_+$:
\begin{align}\label{BTTNIsClosed}
   B_{lmnpq}(\theta_L\Gamma^{lmn}\lambda_L)(\lambda_L\Gamma^{pq}w_+) = 0
\end{align}
Let us prove (\ref{BTTNIsClosed}). This is equivalent to: 
\begin{align}\label{BTFW}
   B_{lmnpq}(\theta_L\Gamma^{lmn}\widehat{\cal F}\;\Gamma^{pq}w_+) = 0
\end{align}
for any self-dual 5-forms ${\cal F}$ and $B$, with $\widehat{\cal F} = {\cal F}_{ijklm}\Gamma^{ijklm}$. To prove (\ref{BTFW}),
we consider particular values for $\widehat{\cal F}$ and $B$. Let us work in the 
{\bf Euclidean signature}: $\Gamma_0^2=\Gamma_1^2=\ldots=1$. Modulo $SO(10)$ rotations, 
there are exactly 3 cases to consider. 
\paragraph     {Case 0}
\begin{align}
   \widehat{\cal F} = \;& \widehat{B} = \Gamma^{01234} + i\Gamma^{56789}
\end{align}
In order to calculate $B_{lmijk}(w_+\Gamma^{lm}\widehat{\cal F}\;\Gamma^{ijk}\theta_L)$, we need:
\begin{align}
\;&   \Gamma^{[01|}(\Gamma^{01234} + i\Gamma^{56789})\Gamma^{|234]} 
+ i\Gamma^{[56|}(\Gamma^{01234} + i\Gamma^{56789})\Gamma^{|789]} \;=
\nonumber\\    
=\;& 
(\Gamma^{01234} + i\Gamma^{56789})^2 = 0
\end{align}
\paragraph     {Case 1}
\begin{align}
   \widehat{\cal F} = \;& \Gamma^{01234} + i\Gamma^{56789}
\\   
\widehat{B} =\;& \Gamma^{01235} - i\Gamma^{46789}
\end{align}
To calculate  $B_{lmijk}(w_+\Gamma^{lm}\widehat{\cal F}\;\Gamma^{ijk}\theta_L)$, consider:
\ifodd\amshow
\begin{align}
   & 120\left(   
   \Gamma^{[01|}(\Gamma^{01234} + i\Gamma^{56789})\Gamma^{|235]} 
-  i\Gamma^{[46|}(\Gamma^{01234} + i\Gamma^{56789})\Gamma^{|789]}  
\right)=
\nonumber \\   
= \;& 72\;\Gamma^{[01|}(\Gamma^{01234} + i\Gamma^{56789})\Gamma^{|23]5} +
48\;\Gamma^{5[0|}(\Gamma^{01234} + i\Gamma^{56789})\Gamma^{|123]} \;-
\nonumber \\   
\;&
- 48\;i\;\Gamma^{4[6|}(\Gamma^{01234} + i\Gamma^{56789})\Gamma^{|789]}  
- 72\;i\;\Gamma^{[67|}(\Gamma^{01234} + i\Gamma^{56789})\Gamma^{|89]4}\;=
\nonumber \\  
=\;&
-72\; \Gamma^{01235}(\Gamma^{01234} - i\Gamma^{56789})
+48\; \Gamma^{01235}(\Gamma^{01234} - i\Gamma^{56789})\;+
\nonumber \\   
\;&
+48\;i\; \Gamma^{46789}(\Gamma^{01234} - i\Gamma^{56789})
-72\;i\; \Gamma^{46789}(\Gamma^{01234} - i\Gamma^{56789})\;=
\nonumber \\   
=\;&
-24\; \Gamma^{01235}(\Gamma^{01234} - i\Gamma^{56789})
-24\;i\; \Gamma^{46789}(\Gamma^{01234} - i\Gamma^{56789})\;=
\nonumber \\  
=\;&24\;\Gamma^{45} +24\;i\; \Gamma^{01236789} -24\;i\;\Gamma^{01236789} -24\;\Gamma^{45} =0
\end{align}
\else
\begin{align}
   & 120\left(   
   \Gamma^{[01|}(\Gamma^{01234} + i\Gamma^{56789})\Gamma^{|235]} 
-  i\Gamma^{[46|}(\Gamma^{01234} + i\Gamma^{56789})\Gamma^{|789]}  
\right)=
\nonumber \\   
= \;& 72\;\Gamma^{[01|}(\Gamma^{01234} + i\Gamma^{56789})\Gamma^{|23]5} +
48\;\Gamma^{5[0|}(\Gamma^{01234} + i\Gamma^{56789})\Gamma^{|123]} \;-
\nonumber \\   
\;&
- 48\;i\;\Gamma^{4[6|}(\Gamma^{01234} + i\Gamma^{56789})\Gamma^{|789]}  
- 72\;i\;\Gamma^{[67|}(\Gamma^{01234} + i\Gamma^{56789})\Gamma^{|89]4}\;=
\nonumber \\  
=\;& 0
\end{align}
\fi
\paragraph     {Case 2}
\begin{align}
   \widehat{\cal F} = \;& \Gamma^{01234} + i\Gamma^{56789}
\\   
\widehat{B} =\;& \Gamma^{01256} + i\Gamma^{34789}
\end{align}
In order to calculate $B_{lmijk}(w_+\Gamma^{lm}\widehat{\cal F}\;\Gamma^{ijk}\theta_L)$, we consider:
\ifodd\amshow
\begin{align}
& 120\left(   
   \Gamma^{[01|}(\Gamma^{01234} + i\Gamma^{56789})\Gamma^{|256]} 
+  i\Gamma^{[34|}(\Gamma^{01234} + i\Gamma^{56789})\Gamma^{|789]}  
\right)=
\nonumber \\  
=\;& 36\; \Gamma^{[01|}(\Gamma^{01234} + i\Gamma^{56789})\Gamma^{|2]56} 
+ 12\;\Gamma^{56}(\Gamma^{01234} + i\Gamma^{56789})\Gamma^{201} \;-
\nonumber \\   
&- 36\;\Gamma^{5[2|}(\Gamma^{01234} + i\Gamma^{56789})\Gamma^{|01]6} 
+ 36\;\Gamma^{6[2|}(\Gamma^{01234} + i\Gamma^{56789})\Gamma^{|01]5} \;+
\nonumber \\   
&+ 12\;i\; \Gamma^{34}(\Gamma^{01234} + i\Gamma^{56789})\Gamma^{789}  
+ 36\;i\;  \Gamma^{[89|}(\Gamma^{01234} + i\Gamma^{56789})\Gamma^{7]34}  \;-
\nonumber \\   
&- 36\;i\;\Gamma^{3[7|}(\Gamma^{01234} + i\Gamma^{56789})\Gamma^{|89]4}  
+ 36\;i\;\Gamma^{4[7|}(\Gamma^{01234} + i\Gamma^{56789})\Gamma^{|89]3}  \;=
\nonumber \\  
=\;&  36\; \Gamma^{01256}(\Gamma^{01234} - i\Gamma^{56789}) 
+ 12\;\Gamma^{01256}(\Gamma^{01234} - i\Gamma^{56789}) \;-
\nonumber \\   
&- 72\;\Gamma^{01256}(\Gamma^{01234} - i\Gamma^{56789})\;-
\nonumber \\   
&- 12\;i\; \Gamma^{34789}(\Gamma^{01234} - i\Gamma^{56789})  
- 36\;i\;  \Gamma^{34789}(\Gamma^{01234} - i\Gamma^{56789}) \;-
\nonumber \\   
&+ 72\;i\;\Gamma^{34789}(\Gamma^{01234} - i\Gamma^{56789}) \; =
\nonumber \\   
=\;& -24 \; \Gamma^{01256}(\Gamma^{01234} - i\Gamma^{56789})
+24 \;i\;\Gamma^{34789}(\Gamma^{01234} - i\Gamma^{56789}) \;=
\nonumber \\   
=\;& 24\;\Gamma^{3456} - 24\;i\;\Gamma^{012789} + 24i\;\Gamma^{012789} - 24\;\Gamma^{3456}=0 
\end{align}
\else
\begin{align}
& 120\left(   
   \Gamma^{[01|}(\Gamma^{01234} + i\Gamma^{56789})\Gamma^{|256]} 
+  i\Gamma^{[34|}(\Gamma^{01234} + i\Gamma^{56789})\Gamma^{|789]}  
\right)=
\nonumber \\  
=\;& 36\; \Gamma^{[01|}(\Gamma^{01234} + i\Gamma^{56789})\Gamma^{|2]56} 
+ 12\;\Gamma^{56}(\Gamma^{01234} + i\Gamma^{56789})\Gamma^{201} \;-
\nonumber \\   
&- 36\;\Gamma^{5[2|}(\Gamma^{01234} + i\Gamma^{56789})\Gamma^{|01]6} 
+ 36\;\Gamma^{6[2|}(\Gamma^{01234} + i\Gamma^{56789})\Gamma^{|01]5} \;+
\nonumber \\   
&+ 12\;i\; \Gamma^{34}(\Gamma^{01234} + i\Gamma^{56789})\Gamma^{789}  
+ 36\;i\;  \Gamma^{[89|}(\Gamma^{01234} + i\Gamma^{56789})\Gamma^{7]34}  \;-
\nonumber \\   
&- 36\;i\;\Gamma^{3[7|}(\Gamma^{01234} + i\Gamma^{56789})\Gamma^{|89]4}  
+ 36\;i\;\Gamma^{4[7|}(\Gamma^{01234} + i\Gamma^{56789})\Gamma^{|89]3}  \;=
\nonumber \\  
=\;& 0 
\end{align}
\fi
Therefore, in this case also  $B_{lmijk}(w_+\Gamma^{lm}\widehat{\cal F}\;\Gamma^{ijk}\theta_L) =0$.
This concludes the proof of (\ref{BTTNIsClosed}).

\paragraph     {Proof that $B_{lmnpq} (\theta\Gamma^{lmn}\theta)\lambda\Gamma^{pq}$ is not  BRST-exact}
The only possibility for it to be BRST-exact would be:
\begin{equation}\label{MaybeExact}
   B_{lmnpq} (\theta_L\Gamma^{lmn}\theta_L)\lambda_L\Gamma^{pq} 
   \stackrel{?}{\simeq}
   Q\left( B_{lmnpq} (\theta_L\Gamma^{lmn}\theta_L)\theta_L\Gamma^{pq} \right)
\end{equation}
The RHS is a linear combination of two BRST-closed expressions:
\begin{equation}
   B_{lmnpq} (\theta_L\Gamma^{lmn}\theta_L) \lambda_L\Gamma^{pq}
\;\mbox{ \tt\small and }\;
   B_{lmnpq} (\theta_L\Gamma^{lmn}\lambda_L) \theta_L\Gamma^{pq}
\end{equation}
These expressions are linearly independent. Indeed, we have:
\begin{align}
   B_{lmnpq} (\theta_L\Gamma^{lmn}\theta_L) 
   (\lambda_L\Gamma^{pq}\;\Gamma^k\lambda_L) =0
\\   
   B_{lmnpq} (\theta_L\Gamma^{lmn}\lambda_L) 
   (\theta_L\Gamma^{pq}\;\Gamma^k\lambda_L)\neq 0
\end{align}
\ifodd\amshow
To prove that $B_{lmnpq} (\theta_L\Gamma^{lmn}\lambda_L) (\theta_L\Gamma^{pq}\;\Gamma^k\lambda_L)\neq 0$, we notice:
\begin{align}
 {\cal A}^k =  B_{lmnpq} (\theta_L\Gamma^{lmn}\lambda_L) (\theta_L\Gamma^{pq}\;\Gamma^k\lambda_L)=
4\;B_{lmnpk}(\theta_L\Gamma^{lmn}\lambda_L)(\lambda_L\Gamma^p\theta_L)
\end{align}
For a fixed $k$ the self-duality of $B$ does not impose any constraint on 
$B_{lmnpk}$, therefore it is enough to prove that:
\begin{equation}
   (\theta_L\Gamma^{[lmn}\lambda_L)(\lambda_L\Gamma^{p]}\theta_L) \neq 0
\end{equation}
For example, consider $\lambda\otimes\lambda=\Gamma^{01234} + i\Gamma^{56789}$, and calculate:
\begin{align}
&  (\theta_L\Gamma^{[012}(\Gamma^{01234} + i\Gamma^{56789})\Gamma^{5]}\theta_L) =
{1\over 2}(\theta_L(\Gamma^{345} - i\Gamma^{0126789})\theta_L) =
\nonumber \\   
= \; & 
(\theta_L\Gamma^{0125}(\Gamma^{01234} - i\Gamma^{56789})\theta_L) \neq 0
\end{align}
\fi
Therefore $\left(Q\left( B_{lmnpq} (\theta_L\Gamma^{lmn}\theta_L)\theta_L\Gamma^{pq} \right)\;\Gamma^k\lambda_L\right)$ is nonzero. \\  
But $  B_{lmnpq} (\theta_L\Gamma^{lmn}\theta_L) 
   (\lambda_L\Gamma^{pq}\;\Gamma^k\lambda_L) $ is zero. 
This implies that (\ref{MaybeExact}) is false.

\subsection{Pure spinor redefinition}
Therefore the vector field $Y_L$ of Section \ref{sec:SpecialCaseB} involves an infinitesimal 
redefinition of the pure spinor field:
\begin{align}\label{YLFor5Form}
   Y_L\lambda_L^{\alpha} = 
B_{lmnpq} (\theta\Gamma^{lmn}\theta)(\lambda\Gamma^{pq})^{\alpha}
\end{align}
which preserves the pure spinor condition: $(\lambda_L\Gamma^kY_L\lambda_L)=0$. 

\section*{Acknowledgments}
We would like to thank N.~Berkovits and O.~Bedoya for discussions and 
comments on the text. This work was 
supported in part by the Ministry of Education and Science of the Russian 
Federation under the project 14.740.11.0347 ``Integrable and 
algebro-geometric structures in string theory and quantum field theory'', 
and in part by the 
RFFI grant 10-02-01315 ``String theory and integrable systems''.

% \bibliographystyle{JHEP} \renewcommand{\refname}{Bibliography}
% \addcontentsline{toc}{section}{Bibliography}
% \bibliography{../andrei}

\def\cprime{$'$} \def\cprime{$'$}
\providecommand{\href}[2]{#2}\begingroup\raggedright\endgroup

\end{document}